\documentclass[lettersize,journal]{IEEEtran}
\usepackage{amsmath,amsfonts}
\usepackage{algorithmic}
\usepackage{algorithm}
\usepackage{array}
\usepackage[caption=false,font=normalsize,labelfont=sf,textfont=sf]{subfig}
\usepackage{textcomp}
\usepackage{stfloats}
\usepackage{url}
\usepackage{verbatim}
\usepackage{graphicx}
\usepackage{cite}
\usepackage{booktabs}
\usepackage{multirow}
\usepackage{tabularx}
\usepackage{color}
\begin{document}

\title{Hierarchical Interpretable Multitask Audio Representation Learning: An Analysis of Event–Scene Relationships through the Information Bottleneck}

\author{Claudia Montero-Ramírez, Alba Martínez-Serrano, Carmen Peláez-Moreno, ~\IEEEmembership{Universidad Carlos III de Madrid}

\thanks{}
\thanks{Manuscript received October 2026}}

\markboth{Journal of \LaTeX\ Class Files,~Vol.~, No.~, October~2026}%
{Shell \MakeLowercase{\textit{et al.}}: A Sample Article Using IEEEtran.cls for IEEE Journals}



\maketitle

\begin{abstract}

Joint Acoustic Scene Classification (ASC) and Audio Tagging (AT) models traditionally rely on unconstrained multihead architectures. However, this design ignores the natural hierarchical dependency between sounds and acoustic scenes. In this work, we investigate a hierarchical (sequential) conditioning paradigm that predicts scene categories directly from intermediate event probabilities. Through a large-scale empirical benchmark with 66 full training runs across three state-of-the-art backbones (PaSST, PANNs, and BEATs) and different loss-weighting settings ($\beta \in [0.0, 1.0]$), we show that hierarchical conditioning achieves competitive classification performance while altering multitask optimization dynamics. Unlike multihead baselines, which sequentially solve the easier ASC task before optimizing AT, the hierarchical architecture enables synchronized dual-task learning for specific loss-weighting settings ($\beta \in [0.6, 0.9]$). Furthermore, we propose a multi-perspective interpretability framework that combines an analysis of the information bottleneck, manifold geometry, and Formal Concept Analysis (FCA) on a Term-Frequency Inverse-Document-Frequency (TF-IDF) event representation, providing complementary insights into event–scene relationships.
Our findings reveal that hierarchical conditioning systematically yields lower conditional entropy, higher mutual information, and superior latent cluster separability (higher Silhouette scores), establishing a more structured and semantically faithful intermediate feature space.
\end{abstract}


\section{Introduction}
\label{sec:introduction}

Sounds are the fundamental component of an acoustic scene. A group of sounds can define an acoustic scene for human understanding. E.g., if we close our eyes and listen to a bird singing, the sound of a waterfall, and the leaves on the trees rustling in the wind, we will surely find ourselves in an outdoor scene, surrounded by nature. 
However, an acoustic scene is rarely defined solely by the isolated sound events occurring within it; rather, human auditory perception relies on both discrete acoustic events and a continuous, ambient background texture—frequently referred to as the acoustic "gist" \cite{oliva2005gist}. In other words, while discrete events (such as footsteps or a dog barking) provide solid semantic reference points, the underlying acoustic gist captures the spatial reverberation and diffuse background noise that characterize the environment globally.

Replicating this ability computationally has become an increasingly important research goal with the rapid development of artificial intelligence (AI). We have known for years that machines can hear—ever since we became capable of recording audio. However, advances in AI have progressively enabled machines to move beyond simply capturing sound towards interpreting what is happening within audio recordings \cite{surrey_cvssp_making_sense}. Along with this paradigm, research communities such as DCASE \cite{dcase} have emerged, launching annual challenges to bring together scientists from around the world to share knowledge in these areas. One of the tasks they propose is Acoustic Scene Classification (ASC) \cite{Schmid2025}, which aims to automatically identify the location where the audio was recorded. They also offer tasks such as Audio Tagging (AT) \cite{Fonseca2019}, which aims to identify which sounds are present in each audio clip, and Sound Event Detection (SED) \cite{WuIACAS2025}, which works in the same way but with the added capability of identifying the specific time points at which those sounds occur. In this work, we focus on AT rather than SED, as our interest lies in the global set of acoustic events that characterize a scene rather than in their temporal occurrence.
A key challenge in studying event-scene relationships is the limited availability of datasets providing both scene and sound event annotations. We therefore focus on \textit{TUT
Sound Events 2016/2017} \cite{heittola2018tut} and \textit{TUT Acoustic Scenes 2016} \cite{mesaros2016tut} and evaluate three pretrained audio backbones— Patchout faSt Spectrogram Transformer (PaSST), Large-Scale Pretrained Audio Neural Networks for Audio Pattern Recognition (PANNs), and Bidirectional Encoder Representation from Audio Transformer (BEATs) — to assess the consistency of our findings across different acoustic representations.

However, going back in time, it was very common to use hand-crafted features as input for models to solve these tasks. These representations included features such as MFCCs, pitch, and the energy of the audio signal, and they enabled predictions to be interpretable. E.g., audio with sustained high energy over time is more likely to come from a noisy place than a quiet one. Over time, AI has become more sophisticated, giving rise to what we call “black boxes”: Deep Learning Networks (DLNs). Alongside them—and more recently—have come transformers, a type of architecture that 
has become absolutely dominant in data-rich settings.
These models achieve fantastic metrics and typically use the spectrogram image as an input feature. However, they are called “black boxes” because, due to their large number of parameters, we lose track of what is happening inside the model from the perspective of interpretability. Nevertheless, techniques have been developed to understand parts of these models, such as SHAP \cite{shap}, LIME \cite{lime}, and the information bottleneck theory \cite{tishby2000information}, proposed by Naftali Tishby.

Beyond understanding what information is used by a model to make a prediction, an additional question arises when several related aspects of the same audio signal are considered: which information is shared across tasks, and which information is specific to each one? This question becomes particularly relevant for acoustic scene understanding, where the environment and the sounds occurring within it are intrinsically related.

\begin{figure}
    \centering
    \includegraphics[width=1\linewidth]{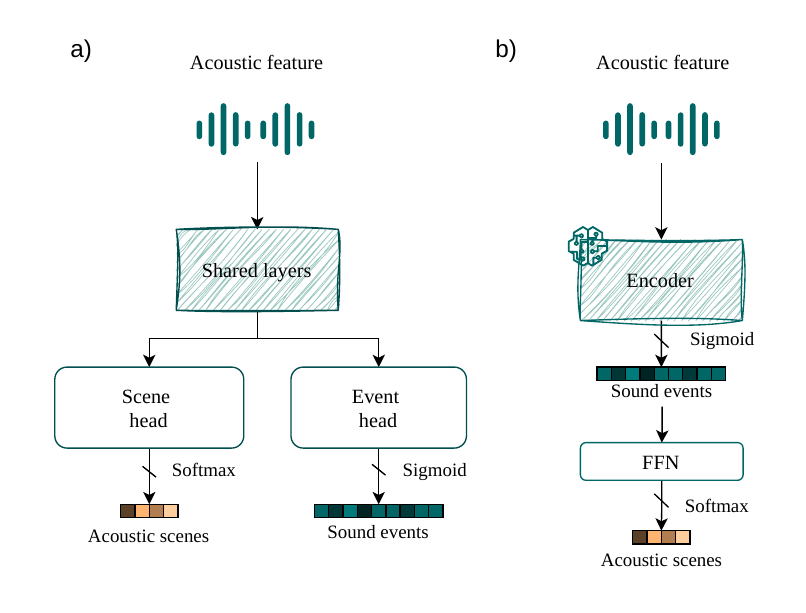}
    \caption{\textbf{a)} Traditional architecture of multitask models with multiple heads \cite{tonami2019, Imoto202, igarashi2023joint}. \textbf{b)} Our proposed hierarchical architecture using an information bottleneck with a feed-forward network (FFN).}
    \label{fig:compare_architectures}
\end{figure}

A promising strategy for exploiting these relationships is multitask learning. In this paradigm, a single model is trained to perform several tasks simultaneously, sharing part of its internal representations. The idea is that related tasks can benefit from one another, since the information learned in one task acts as an inductive bias for the others. In the context of audio, this allows a model to simultaneously learn to classify acoustic scenes (ASC) and detect sounds or events (AT/SED) using a shared backbone that extracts general representations from the audio data.
After the backbone, multihead layers are added so that there are small, specialized models for each task, as shown in Fig. \ref{fig:compare_architectures}.

Most existing multitask audio models adopt a multihead architecture, in which task-specific prediction heads operate on a shared representation learned by a common backbone. While this design enables the joint learning of related tasks such as ASC and AT, it does not explicitly constrain how information about acoustic events contributes to the representation used for scene prediction. This raises an important question: \textbf{how are event-level acoustic concepts preserved and organized when learning to predict the scene, and how does this organization depend on the architecture used for multitask learning?}

Despite the growing interest in multitask audio representation learning, most existing approaches focus primarily on predictive performance, while comparatively little attention has been devoted to understanding how acoustic event information is organized within representations that also support scene prediction. In particular, it remains unclear which event information is preserved, how strongly it is associated with the predicted scene, and whether different multitask architectures induce different event-scene relationships. Addressing these questions requires going beyond classification performance and examining the learned representations themselves.

This motivates us to introduce a hierarchical architecture in which the intermediate event representation directly conditions the scene prediction. Rather than treating event and scene predictions as parallel outputs of a shared representation, this formulation explicitly models the event-to-scene dependency and introduces a well-defined intermediate bottleneck through which this relationship can be examined (Fig. \ref{fig:compare_architectures}). The resulting architecture therefore provides not only an alternative way of performing joint ASC and AT, but also a natural framework for studying how event information is preserved and transformed when predicting acoustic scenes.



The main objective of this work is therefore to investigate how the choice of multitask architecture affects the relationship between acoustic events and acoustic scenes in pretrained audio models. To this end, we compare conventional multihead multitask learning with a hierarchical architecture that uses the intermediate event representation to predict the acoustic scene.

More specifically, this work aims to:

\begin{enumerate}
    \item Evaluate the hierarchical and multihead architectures across different pretrained audio backbones and task-loss weights, analyzing their effect on ASC and AT performance.
    
    \item Characterize the event representations learned by both architectures using information-theoretic and geometric measures, including conditional entropy, mutual information, KL divergence, and clustering quality.
    
    \item Analyze how the learned event representations preserve the semantic relationships between events and scenes, using Term-Frequency Inverse-Document-Frequency (TF-IDF) weighting and Formal Concept Analysis (FCA).
    
    \item Assess whether the explicit event-to-scene bottleneck provides a more transparent representation of event-scene relationships than the conventional multihead architecture.
\end{enumerate}

Based on these objectives, the main contributions of this work are:

\begin{itemize}

    \item \textbf{Hierarchical multitask architecture for event--scene modeling:} We introduce a hierarchical alternative to the conventional multihead architecture for joint ASC-AT learning, in which the intermediate event representation is explicitly used to condition the scene prediction. This architecture provides a direct pathway for analyzing the relationship between acoustic events and scenes.

   \item \textbf{Comprehensive large-scale benchmark:} Rather than introducing isolated models, we conduct an extensive empirical benchmark comprising 66 full training runs across three state-of-the-art pretrained backbones (PaSST, PANNs, and BEATs) under exhaustive loss-weighting regimes ($\beta \in [0.0, 1.0]$), providing the first systematic study of architectural topology in joint ASC-AT tasks.

    \item \textbf{Synchronous multitask learning dynamics:} We find a fundamental difference in optimization behavior on this dataset: while multihead models sequentially master the easier ASC task before optimizing AT, the hierarchical paradigm achieves synchronized dual-task learning across balanced loss weights ($\beta \in [0.6, 0.9]$), leading to more stable joint representations.

    \item \textbf{Multi-perspective analysis of event-scene representations}: We introduce a multi-perspective interpretability framework combining information bottleneck metrics ($H(E|S)$, $I(E; S)$), manifold geometry (Silhouette Score over $P(E|S)$), and FCA on TF-IDF weighted event matrices to evaluate how discrete acoustic concepts are preserved through the intermediate bottleneck.
\end{itemize}

\section{Related Work}
\label{section:related_work}


ASC, SED, and AT have been extensively studied within the DCASE community \cite{dcase}. In recent years, ASC research has increasingly focused not only on classification performance but also on computational efficiency. In the 2023 DCASE challenge, \textit{Schmid et al.} \cite{Schmid2023} proposed a knowledge-distillation approach that achieved the best performance in the challenge, using PaSST \cite{PaSST} as teacher model. PaSST is an efficient and high-capacity transformer architecture that has subsequently become widely adopted for acoustic scene analysis \cite{monterobuilding2026}. In SED and AT, large-scale pretrained audio encoders such as PANNs \cite{panns}, trained on AudioSet \cite{AudioSet}, have become standard architectures for learning general-purpose acoustic representations. Their representations have also been incorporated into more complex multimodal and audio-language systems, such as Large-Scale Contrastive Language-Audio Pretraining with Feature Fusion and Keyword-to-Caption Augmentation (CLAP) \cite{clap}.

Regarding databases, one of the most well-known in this field is AudioSet \cite{AudioSet}, which contains human-labeled 10-second sound clips extracted from 2, 084, 320 YouTube videos. There are also other databases widely used for AT and SED, such as FSD50K \cite{fonseca2022FSD50K} or DESED \cite{desed2019}; this last one is specialized in domestic sounds and has been used since DCASE 2020. Similarly, in 2024, WildDESED \cite{Xiao2024WildDESED} was created; an extension of the original DESED dataset to reflect a wider variety of domestic scenarios by incorporating complex and unpredictable background noises.
On the other hand, the TAU Urban Acoustic Scenes 2022 Mobile Development Data Set \cite{heittola_2022_6337421} was selected for the ASC task since 2022. Similarly, other ASC datasets such as TUT Urban Acoustic Scenes 2018 \cite{heittola2018tut} have been used for years in the DCASE challenge.

Despite the close complementarity between ASC, SED, and AT, these tasks have traditionally been investigated independently, overlooking the potentially informative relationships between acoustic events and the environments in which they occur. This has motivated the application of multitask learning to jointly exploit scene- and event-level information. \textit{Tonami et al.} \cite{tonami2019} investigated joint ASC and event recognition as a multitask problem, providing an early baseline for learning both types of information from shared acoustic representations. To this end, they adapted the TUT Sound Events 2016/2017  and TUT Acoustic Scenes 2016 \cite{mesaros2017dcase}, \cite{mesaros2016tut} datasets to this joint learning setting. 

Subsequent studies also adopted this baseline and explored different mechanisms for exploiting the interaction between the tasks. \textit{Imoto et al.} \cite{Imoto202} introduced soft scene labels generated by a teacher model to improve joint learning, while \textit{Igarashi et al.} \cite{igarashi2023joint} investigated semi-supervised joint learning by incorporating weak event labels. Other approaches have explored additional sources of information, such as semantic scene context \cite{tonami2022semantic}, adversarial training \cite{igarashi2022information}, balanced multitask objectives \cite{Zhang2024}, and scene-dependent representations \cite{zhang2025scene}. More broadly, joint learning of ASC, SED, and AT has also been investigated within the DCASE community, demonstrating the potential of multitask representations to capture complementary information across acoustic tasks.

A particularly relevant line of research explicitly models the relationship between acoustic events and scenes. Hou et al. \cite{Hou2022} proposed Relation-Guided Acoustic Scene Classification (RGASC), in which event embeddings obtained from PANNs are used to construct an event--scene relation matrix that guides scene classification. This idea was further developed by Hou et al. \cite{Hou2024} using additional pretraining on the JSSED synthetic dataset. These approaches demonstrate that event-level information can provide useful contextual cues for ASC beyond conventional shared representations. However, their primary objective is to use event information to improve scene classification, rather than to characterize how the learned representation itself encodes the underlying event--scene relationships.

There is a great deal of research focused on training neural networks through an analysis based on information theory. One of the best-known researchers in this field is Naftali Tishby, who proposed the \textit{Information Bottleneck} in 2000~\cite{tishby2000information}. This is a theoretical principle in machine learning and information theory, and the main idea is that learning can be understood as a process of compressing information. The goal is to find a representation of the $X$ data that retains only the information relevant to predicting a target variable $Y$, while discarding irrelevant information. 
In 2019, Tishby explained his groundbreaking proposal at the \textit{Conference on the Mathematical Theory of Deep Learning (DeepMath 2019)}, where he described the two phases of neural network training~\cite{tishby2019deepmath}. Letting $T$ be a latent variable, he identified a first \textit{fitting phase} in which the network memorizes the input information, increasing both $I(X;T)$ and $I(T;Y)$ simultaneously. Afterwards, there is a second \textit{compression phase} where the network discards irrelevant information and generalizes, reducing $I(X;T)$ while $I(T;Y)$ remains at a similar level.
Moreover, deeper architectures tend to reach the compression regime faster, as multiple layers facilitate hierarchical abstraction \cite{tishby2017dss}. 

Following the analysis of relationships between acoustic events and scenes, FCA provides a natural formal framework for representing object--attribute relations through concept lattices, making their underlying structure explicit \cite{ganter1999formal, valverde2006towards}.
Additionally, beyond joint deep representations, techniques from Information Retrieval (IR) offer an explicit framework for measuring the occurrence and relevance of discrete acoustic elements within larger acoustic environments. In particular, the TF-IDF \cite{sparck1972statistical} formulation—traditionally used in text processing—has been adapted to audio analysis by modeling acoustic scenes as "documents" and detected sound events or acoustic words as "terms" \cite{montero2024spatio, Luis-Mingueza2022BridgingApproach}. Frameworks based on Bag-of-Audio-Words (BoAW) \cite{pancoast2012bag} and event-based semantic vectors use TF-IDF to reduce the importance of common, uninformative background sounds while emphasizing distinctive events that strongly characterize a specific scene \cite{Luis-Mingueza2022BridgingApproach}. In a sequential audio retrieval and classification paradigm—where sound events are predicted before scene inference—TF-IDF provides an interpretable intermediate layer. It quantifies how informative the predicted event sequence is about the underlying scene, establishing an explicit information-theoretic bridge to evaluate how effectively discrete event representations transmit the necessary context for downstream scene analysis.

Despite the continuous progress in multitask audio representation learning, no prior work has proposed a strictly hierarchical, sequential framework that analyzes the bottleneck of acoustic event representations through the combined approaches of Information Theory and Information Retrieval. Most existing literature focuses primarily on proposing novel multi-head or joint-learning architectures aimed at maximizing classification metrics. Consequently, fundamental research on how acoustic events and scenes interact within a single representation is lacking, particularly on datasets of this nature. While \textit{Hou et al.} \cite{Hou2022} introduced RGASC to model event-scene dependencies via a relation matrix constructed from PANN embeddings, their approach uses event information only as a contextual feature, not as a strict sequential bottleneck. 

In this work, we address this gap by introducing a hierarchical multitask architecture in which the intermediate event representation explicitly conditions the scene prediction, and by systematically comparing it with the conventional multihead architecture across multiple pretrained audio backbones and loss-weighting configurations. Beyond predictive performance, we analyze the resulting representations from complementary information-theoretic, geometric, and semantic perspectives, combining entropy and mutual information measures, representation geometry, and TF-IDF-based FCA to examine how event-scene relationships are preserved through the learned bottleneck.

\color{black}


\section{Dataset Analysis via FCA lattices}

\subsection{Dataset Overview}

\begin{figure}[H]
    \centering    
    \includegraphics[width=1\linewidth]{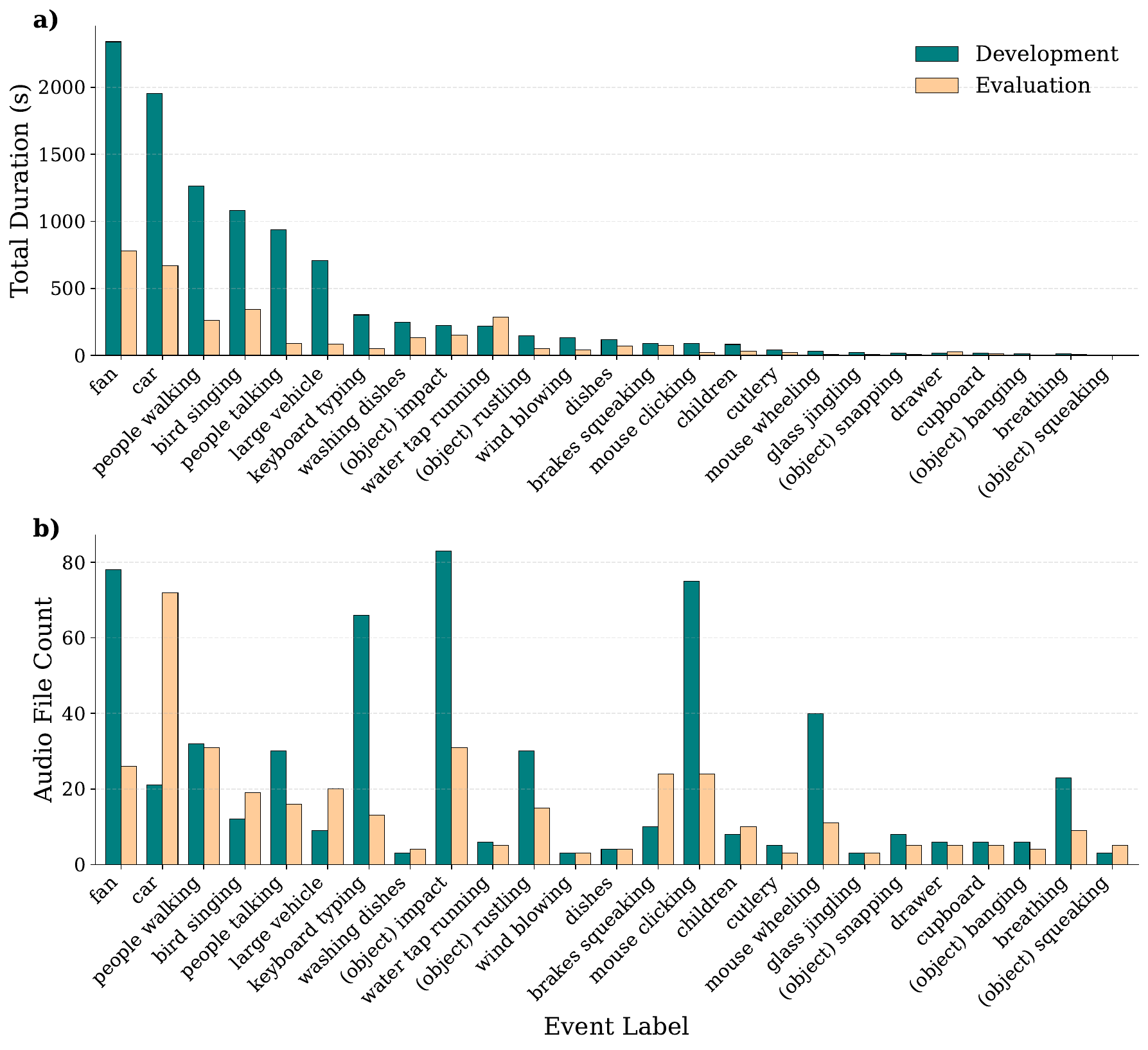}
    \caption{Sound Event distribution for development and evaluation sets. \textbf{a)} Distribution regarding the total duration (in seconds) of each sound event label. \textbf{b)}Frequency of occurrence for each sound event based on unique audio file counts. Teal bars represent the development set, while light orange bars represent the evaluation set. All labels are sorted according to the development set duration for consistency.}
    \label{fig:event_distribution}
\end{figure}

\begin{figure}[H]
    \centering
    \includegraphics[width=0.7\linewidth]{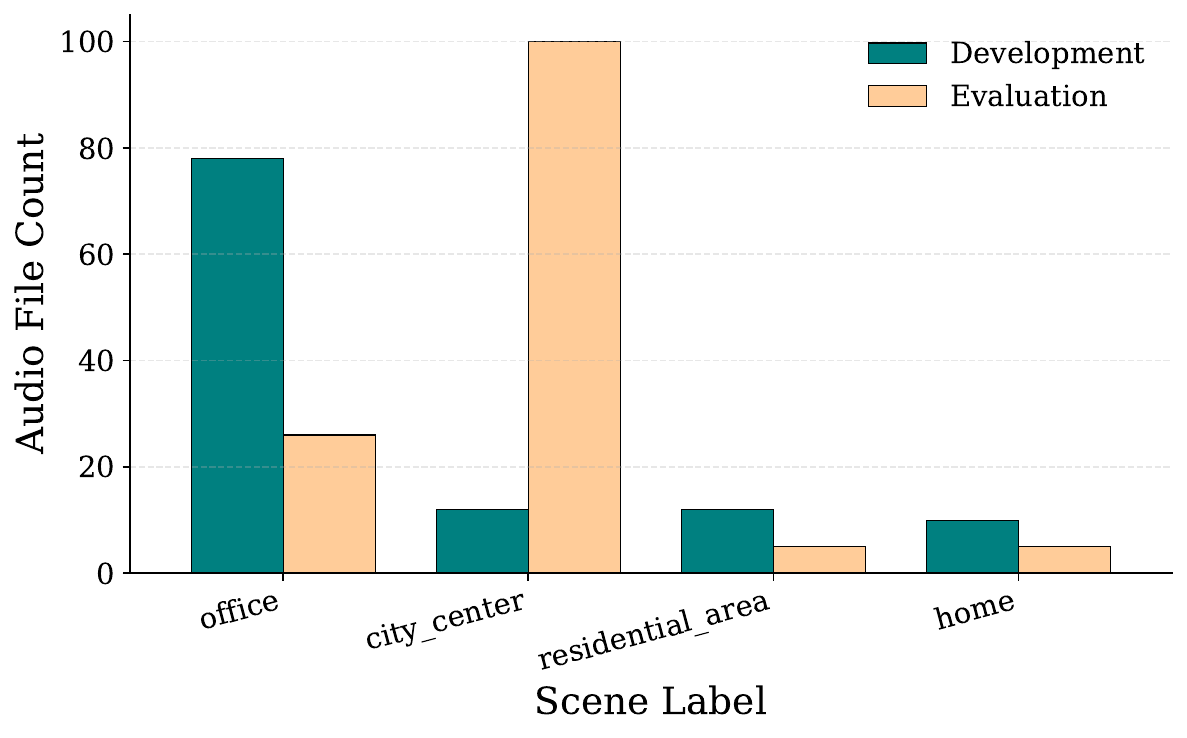}
    \caption{Acoustic scene distribution for development and evaluation sets. The bars represent the total number of unique audio recordings associated with each scene label. Teal and light orange bars denote the development and evaluation partitions, respectively.}
    \label{fig:scene_distribution}
\end{figure}

To perform this study, we have used the TUT Sound Events 2016/2017 and TUT Acoustic Scenes 2016 datasets \cite{mesaros2016tut, mesaros2017dcase}. The data has already been split into two sets: a development set and an evaluation set. Each of these contains $25$ sound labels and $4$ acoustic scene labels. The distribution and the names of labels can be shown in Fig. \ref{fig:event_distribution} for sound events and in Fig. \ref{fig:scene_distribution} for acoustic scenes.

In total, there are $192$ min of recordings in the development set and $74$ min in the evaluation set.

\subsection{Event-Scene Relationship Analysis via FCA Lattices}
\label{sec:dataset:FCA}

To better understand the relationship between sound events and acoustic scenes, we employ FCA as a complementary analysis of the dataset. FCA provides a mathematical framework to identify groups of attributes that tend to occur together across a set of objects. In our case, each acoustic scene is considered an object, while the sound-event labels are treated as attributes. A binary formal context is constructed by indicating whether a given sound event accounts for a sufficient proportion of the scene duration. This representation allows us to identify recurring combinations of sound events and to characterize the acoustic composition of the different scenes.

\begin{figure*}
    \centering
    \includegraphics[width=1\textwidth]{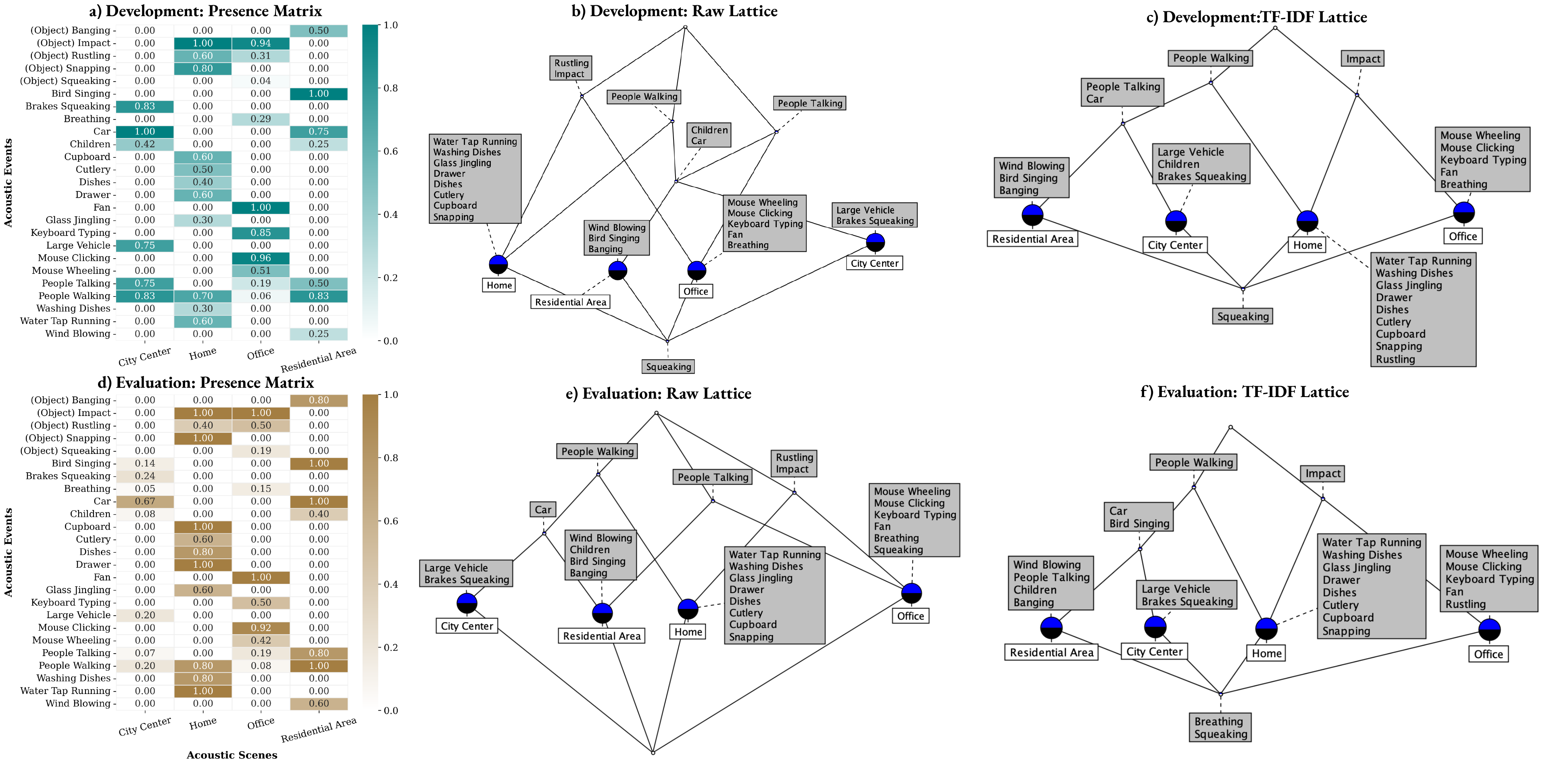}
    \caption{Event-scene relationships for the development (top row) and evaluation (bottom row) subsets. \textbf{(a,d)} Scene-event presence matrices, where each entry is the proportion of the scene duration in which the event is active. \textbf{(b,e)} FCA lattices of the formal contexts obtained by binarizing these matrices with a threshold of $\varphi=0.15$. \textbf{(c,f)} FCA lattices obtained by applying the same threshold after TF-IDF weighting (smoothed IDF and per-scene L2 normalization), which down-weights events shared across scenes and highlights discriminative event-scene associations. In each lattice, blue nodes denote scenes (objects) and grey labels denote events (attributes); an event is associated with every scene reachable downwards from its label, and events at the bottom node are not associated with any scene.}
    \label{fig:fca_dataset}
\end{figure*}

For each acoustic scene, we compute a sound presence matrix describing the relative temporal prevalence of each sound event. The total duration associated with each event is obtained by summing its annotated durations across all audio recordings belonging to a given scene. These values are then normalized by the total duration of the corresponding scene, resulting in a scene--event matrix whose entries represent the proportion of the scene duration associated with each event. This matrix is used to construct the binary formal contexts by applying a threshold $\varphi$; after inspecting the resulting prior distributions, we set $\varphi = 0.15$ to obtain the formal contexts used for the FCA analysis. Fig. \ref{fig:fca_dataset} shows the sound presence matrix, the lattice applied to it, and the lattice applied to the TF-IDF transformation of the prior matrix, for the development and evaluation sets. 
To obtain a more discriminative view, we additionally apply a TF-IDF transformation to the scene–event matrix, treating each scene as a document and each event as a term whose frequency is given by its relative temporal prevalence. The inverse document frequency factor down-weights events that occur across many scenes (e.g., "people walking", present in three of the four scenes), while the subsequent per-scene L2 normalization expresses each weight relative to the overall event profile of that scene. Consequently, an event only exceeds $\varphi$ when it is both prevalent within a scene and specific to it, so the same threshold selects discriminative event–scene associations rather than merely frequent ones. Note that, because of this normalization, TF-IDF can also promote weak but specific events in scenes with sparse event profiles, and demote secondary events in scenes with many dominant events.

In Fig. 4, we see that this dataset contains many events that are highly representative of a particular acoustic scene. For example, both "breathing" and "fan" appear exclusively in the "office" scene in the development set. More generally, the raw lattices (Fig. \ref{fig:fca_dataset} \textbf{(b,e)}) show that each scene is characterized by a block of exclusive events attached directly to its object concept: domestic activity sounds ("water tap running", "washing dishes", "glass jingling", "drawer", "dishes", "cutlery", "cupboard", "snapping") for "home"; computer-related sounds ("keyboard typing", "mouse clicking", "mouse wheeling") together with "fan" and "breathing" for "office"; traffic sounds ("large vehicle", "brakes squeaking") for "city center"; and natural or outdoor sounds ("wind blowing", "bird singing", "banging") for "residential area".

Above these scene-specific concepts, the lattices reveal a small number of shared concepts that capture the overlap between scenes. The events "impact" and "rustling" group "home" and "office" into an indoor concept related to object manipulation, whereas "car" (and "children" in the development set) groups "city center" and "residential area" into an outdoor concept. "People walking" is the most widespread event, linking "home" with both outdoor scenes, while "people talking" links "office" with the outdoor scenes. The latter creates crossing relationships between the indoor and outdoor branches, so the raw lattices (10 concepts each) do not follow a clean hierarchical structure. The outdoor scenes share the largest number of events (e.g., four of the six events of "city center" in the development set are also present in "residential area"), which suggests that this pair is the most acoustically confusable.

The TF-IDF lattices (Fig. \ref{fig:fca_dataset} \textbf{(c,f)}) simplify this structure to 9 concepts. Since "people talking" is no longer associated with "office", the indoor and outdoor branches become clearly separated: "home" and "office" are joined only by "impact", "city center" and "residential area" by "car" (together with "people talking" in the development set and "bird singing" in the evaluation set), and "people walking" remains the only bridge between "home" and the outdoor scenes. In addition, events that were shared by two scenes are assigned to the scene where they are more prevalent. For instance, "children" becomes exclusive to "city center" in the development set, and "rustling" becomes exclusive to "home" in the development set but to "office" in the evaluation set. This last case illustrates that some events are genuinely ambiguous between scenes, and that their discriminative value depends on the subset considered.

Comparing both subsets, the core structure is stable: the exclusive event blocks of each scene, the outdoor concept defined by "car", the indoor concept defined by "impact", and the bridging role of "people walking" appear in all four lattices. The differences are concentrated in low-prevalence events whose values lie close to $\varphi$, such as "children", "squeaking", and "breathing". "Squeaking", for example, is not associated with any scene in the development set but appears in "office" in the raw evaluation lattice, while "breathing" and "squeaking" fall below the threshold after the TF-IDF transformation in the evaluation set. Conversely, "bird singing" reaches "city center" in the evaluation set only after TF-IDF, owing to the sparse event profile of this scene. Overall, the lattices show that acoustic scenes are largely defined by specific combinations of sound events organized around a few shared concepts, which supports conditioning scene predictions on intermediate event probabilities, as in the proposed hierarchical architecture.

\section{Proposed Method}
\label{section_methods}

\subsection{Model architectures}

\begin{figure}
    \centering
    \includegraphics[width=1\linewidth]{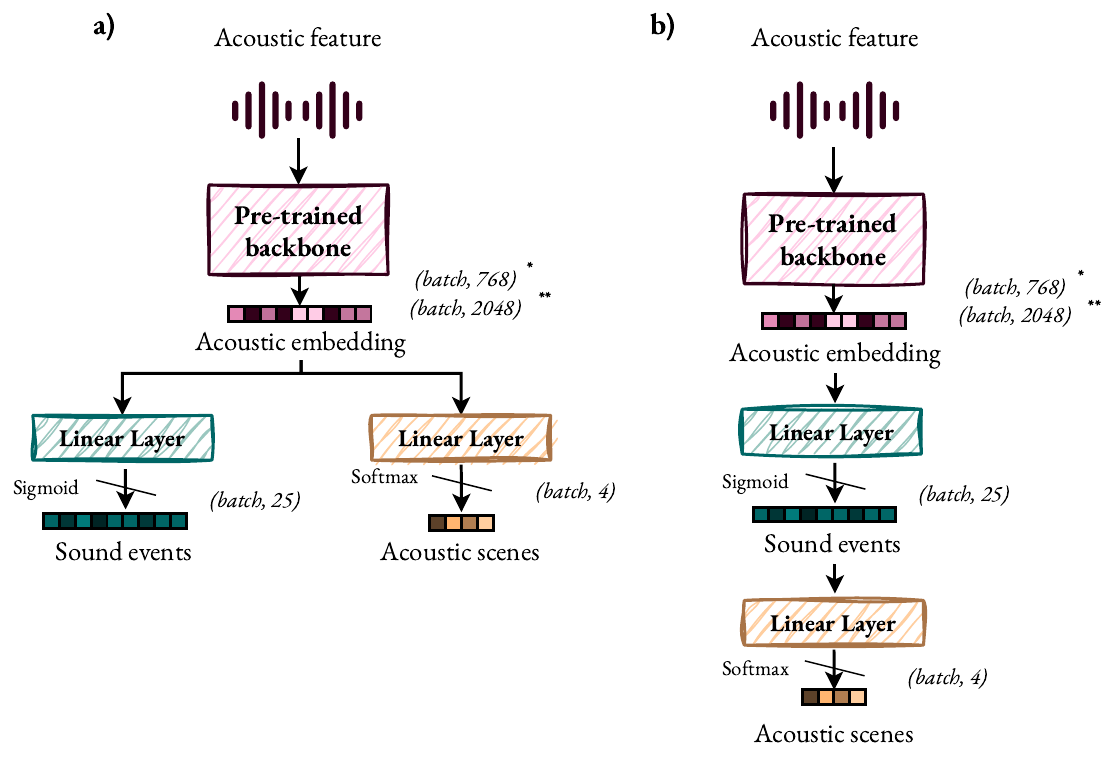}
    \caption{Exact network configuration of the models with \textbf{a)} multihead and \textbf{b)} hierarchical architectures. \textbf{*} BEATs and PaSST use 768-dimensional acoustic embeddings, whereas \textbf{**} PANNs uses 2048-dimensional embeddings.}
    \label{fig:networks}
\end{figure}

We use three models that are widely recognized in the scientific community, and all of them were pre-trained on AudioSet. We trained every model using two different configurations: multihead and hierarchical architectures, as illustrated in Fig. \ref{fig:networks}.

\subsubsection{\textbf{PaSST}}
In recent years, transformer models have consistently outperformed traditional architectures across a variety of tasks. Inspired by their success in natural language processing, Vision Transformers (ViTs) \cite{ViT} have been introduced for image analysis, demonstrating strong performance and rapidly gaining popularity in computer vision applications.
PaSST is ViT applied to spectrogram images \cite{PaSST}. The winners of DCASE Task 1 in 2023, \textit{Schmid et al.} \cite{Schmid2023}, used it as a teacher model to distill knowledge for CP-Mobile.
The most notable feature of this model is the use of the Patchout technique. Not only does it achieve very high performance across all the aforementioned tasks, but it does so efficiently, making it suitable for running experiments on any single consumer-grade GPU. 

\begin{figure}
    \centering
    \includegraphics[width=1\linewidth]{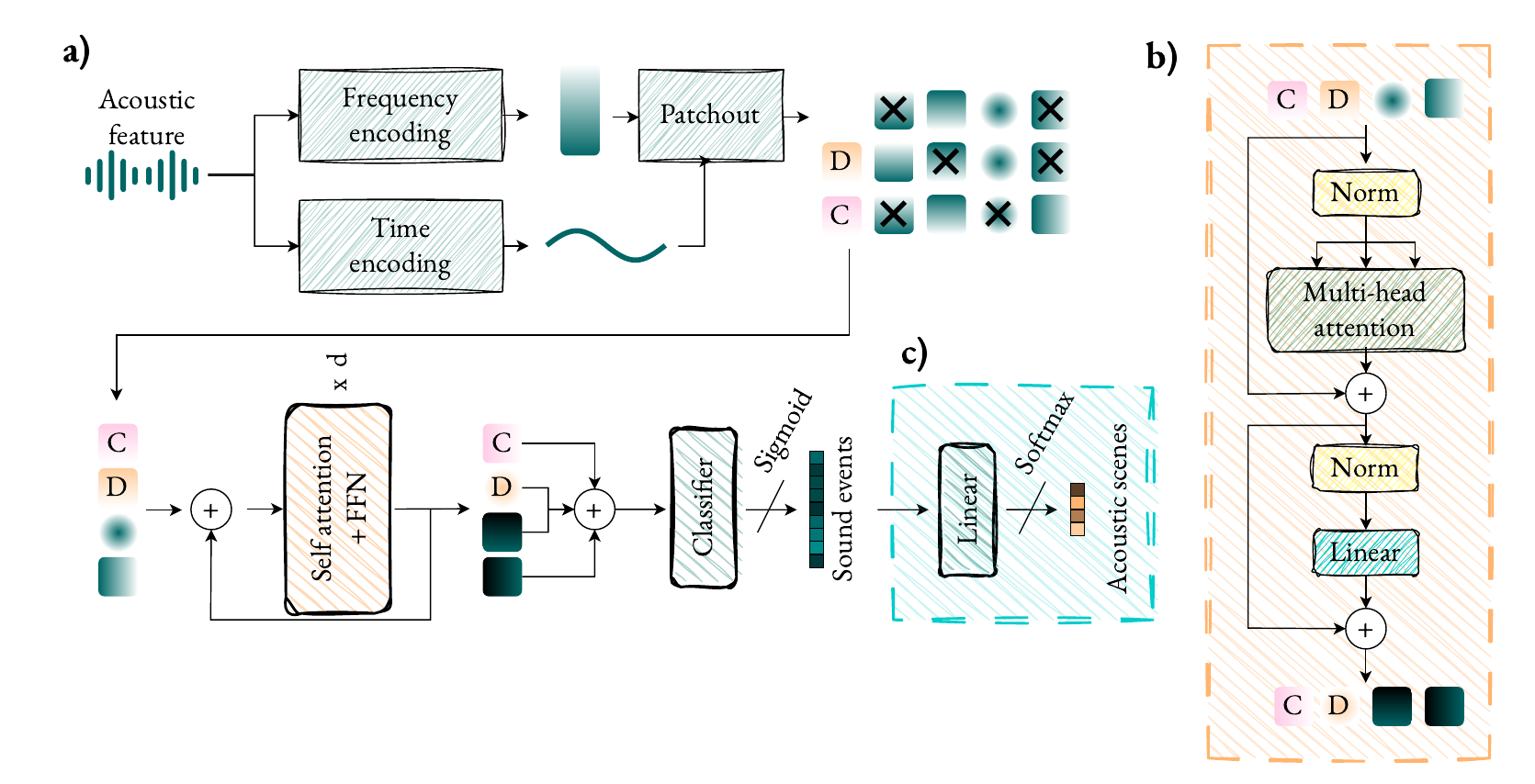}
    \caption{The Patchout transformer (PaSST) adapted to a hierarchical architecture: (\textbf{a}) Overall architecture; C: classification token and D: distillation token \cite{PaSST}. (\textbf{b}) The architecture of the Self-attention layer + FFN (Feedforward network) \cite{ViT}. \textbf{(c)} Hierarchical extension of the model to predict acoustic scenes from sound events.} 
    \label{fig:passt_architecture}
\end{figure}

PaSST was pre-trained for SED \cite{PaSST} on AudioSet, a dataset that contains $2.1$ million annotated videos, $5.8$ thousand hours of audio, and $527$ classes of annotated sounds. From this starting point, we performed hierarchical transfer learning by replacing the final layer of $527$ outputs with one of $25$ corresponding to the sound labels of the TUT Sound Events 2016/2017 and TUT Acoustic Scenes 2016 datasets. Afterwards, we added a Feedforward Network formed by a single Linear Layer to predict $4$ acoustic scenes. The overall architecture can be shown in Fig. \ref{fig:passt_architecture}, where \textbf{(a)} and \textbf{(b)} correspond to the original PaSST architecture, and \textbf{(c)} represents our contribution related to hierarchical transfer learning with a sequential architecture. 

\subsubsection{\textbf{BEATs}}

Similar to PaSST, BEATs (Bidirectional Encoder representation from Audio Transformers) employs a ViT architecture as its backbone, using a linear projection layer followed by a stack of Transformer encoder layers to process sequences of audio patches. In contrast to conventional audio pre-training approaches based on acoustic feature reconstruction, BEATs is pre-trained through a Masked Audio Modeling (MAM) objective, where masked audio patches are used to predict discrete acoustic tokens generated by an iterative acoustic tokenizer. The acoustic tokenizer and the audio SSL model are iteratively optimized through knowledge distillation, allowing both components to improve their representations progressively. They then fine-tune the resulting audio representation model for downstream audio classification tasks by replacing the pre-training label predictor with a task-specific linear classifier \cite{chen2022beatsaudiopretrainingacoustic}. In this work, we use the BEATs model pre-trained on AudioSet as a checkpoint and replace its final linear classification layer with two alternative architectures: a hierarchical sequential architecture and a multi-head architecture, following the same strategy used for the other backbone models.

\subsubsection{\textbf{PANNs}}

These models constitute one of the most widely used families of neural networks for general-purpose audio classification and sound event recognition. Although introduced before the growth of attention and Transformer-based architectures, PANNs models remain highly competitive and continue to be widely adopted as audio encoders in more complex multimodal and audio-language systems, such as CLAP \cite{clap}. Their strong performance, robustness across different audio domains, and availability of pre-trained models make PANNs a relevant convolutional baseline for comparison with more recent Transformer-based architectures. In this work, we use a pre-trained PANNs model as one of the backbone architectures and adapt its final classification layer to the two proposed configurations: the hierarchical sequential architecture and the multi-head architecture.
\subsection{Joint Training and Loss Formulation}

Given the multitask nature of the problem, a composite loss function is employed, as is done in the literature \cite{zhang2025scene}. This is defined in Eq.~\ref{eq:loss}.

\begin{equation} \label{eq:loss}
    \mathcal{L} = \beta \cdot \mathcal{L_{\text{event}}} + (1-\beta) \cdot \mathcal{L_{\text{scene}}}
\end{equation}

where $\beta$ is a weighting parameter that controls the contribution of each loss term. It was selected through cross-validation by varying its value from $0$ to $1$ with a step size of $0.1$.

On the other hand, $\mathcal{L}_{\text{event}}$ corresponds to the multi-label acoustic event detection loss, defined as the Binary Cross-Entropy (BCE) loss across all audio segments and event classes (Eq.~\ref{eq:loss_event}):

\begin{equation}
\label{eq:loss_event}
\begin{aligned}
\mathcal{L}_{\text{event}} = -\frac{1}{N \cdot |E|} \sum_{n=1}^{N} \sum_{e \in E} \Big[&
y_{n,e} \log(\hat{p}_{n,e}) \\
&+ (1 - y_{n,e}) \log(1 - \hat{p}_{n,e})
\Big]
\end{aligned}
\end{equation}

where $N$ is the total number of audio segments, $E$ is the set of acoustic event classes, $y_{n,e} \in \{0, 1\}$ represents the ground-truth binary label indicating the presence of event $e$ in segment $n$, and $\hat{p}_{n,e} = \sigma(t_{n,e}) \in [0, 1]$ is the predicted probability for event $e$, computed by applying the sigmoid function $\sigma(\cdot)$ to the model's logit output $t_{n,e}$. This formulation corresponds to the Binary Cross-Entropy Loss\footnote{\url{https://docs.pytorch.org/docs/stable/generated/torch.nn.BCEWithLogitsLoss.html}}.

Finally, for Acoustic Scene Classification (ASC), we use the Cross-Entropy loss across all segments and scene classes, defined in Eq.~\ref{eq:loss_scene}:

\begin{equation} 
\label{eq:loss_scene}
\mathcal{L}_{\text{scene}} = -\frac{1}{N} \sum_{n=1}^{N} \sum_{s \in S} y_{n,s} \log(\hat{q}_{n,s})
\end{equation}

where $S$ denotes the set of acoustic scene classes, $y_{n,s} \in \{0, 1\}$ is an indicator variable specifying whether ground-truth scene $s$ is assigned to segment $n$ (one-hot encoded), and $\hat{q}_{n,s} \in [0, 1]$ is the predicted probability for scene $s$ on segment $n$, obtained via the softmax activation over the scene output logits. This corresponds to the standard Cross-Entropy Loss\footnote{\url{https://docs.pytorch.org/docs/stable/generated/torch.nn.CrossEntropyLoss.html}}.

\begin{figure}
    \centering
    \includegraphics[width=1\linewidth]{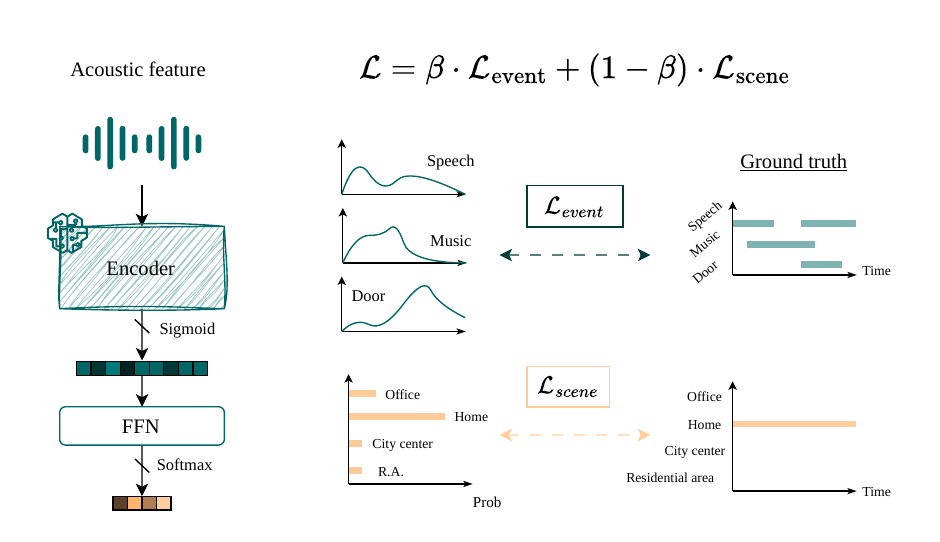}
    \caption{Summary diagram of the combined loss calculation.}
    \label{fig:our_model}
\end{figure}

Fig. \ref{fig:our_model} shows a summary diagram of the loss calculation. 

\subsection{Information-Theoretic Framework} \label{sec:info_framework}

According to Naftali Tishby's theory about the information bottleneck \cite{tishby2000information}, learning can be understood as a process of compressing information. Given that $X$ represents the input data, the goal is to find a latent variable $T$ that retains only the information relevant to predict a target variable $Y$. In $2019$, Thisby proposed the two phases of neural network training at the \textit{Conference on the Mathematical Theory of Deep Learning (DeepMath 2019)} \cite{tishby2019deepmath}. He identified a first \textit{fitting phase} in which the network memorizes the input information, increasing both $I(X;T)$ and $I(T;Y)$ simultaneously. Afterwards, there is a second \textit{compression phase} where the network discards irrelevant information and generalizes, reducing $I(X;T)$ while $I(T;Y)$ remains at a similar level.

In our multitask formulation, we adapt this framework by forcing the intermediate latent representation $T$ to explicitly represent the predicted acoustic event space $E$, while the target variable $Y$ corresponds to the acoustic scene $S$. The input audio signal $X$ is processed by the deep backbone until it is compressed into this intermediate event-level probability space $E$. This event space acts as the functional \textit{information bottleneck} of our system: the network must discard noisy acoustic variability from $X$ while retaining only the sound event descriptors necessary to predict the scene $S$.

\begin{figure}[H]
    \centering
    \includegraphics[width=1\linewidth]{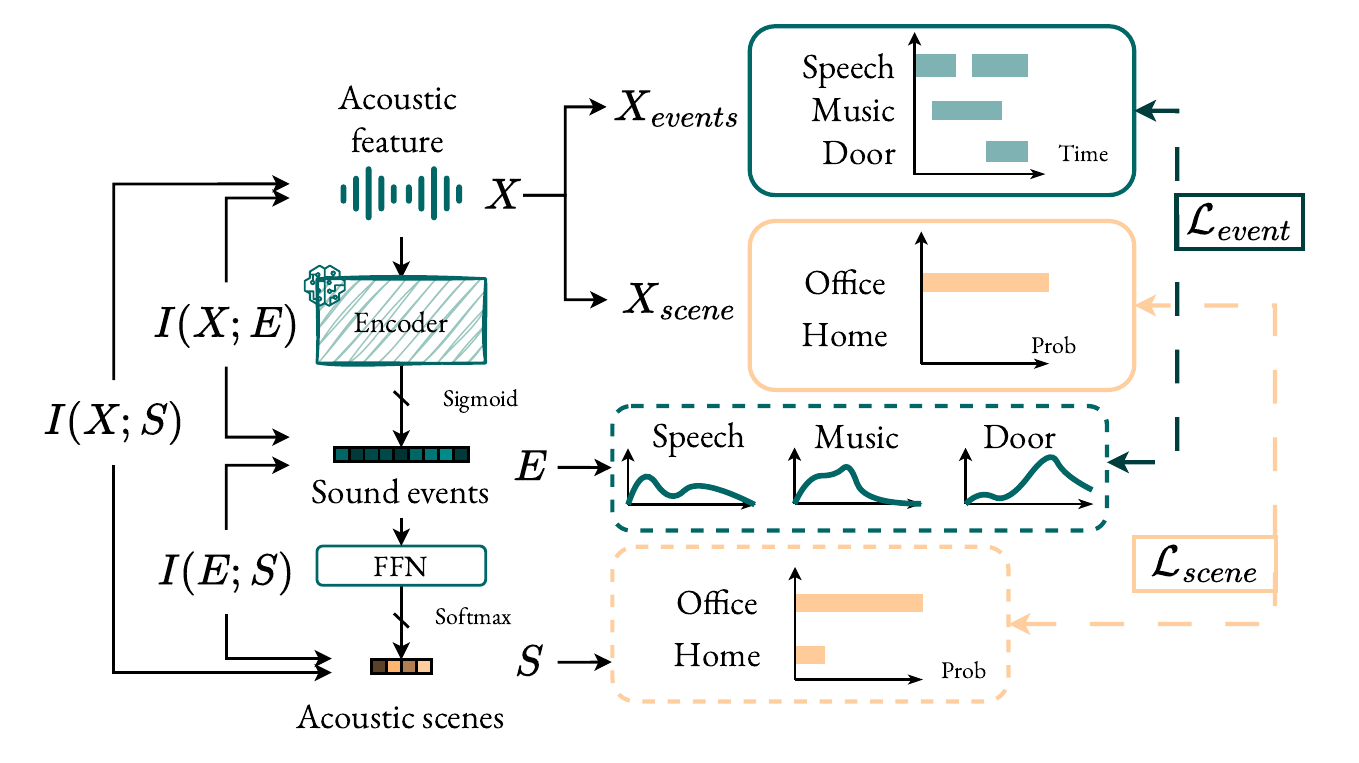}
    \caption{Overview of the proposed hierarchical information bottleneck framework, where the acoustic input $X$, comprising the data and ground truth labels $X_{\mathrm{gt}} = \{{X_{\mathrm{events}}, X_{\mathrm{scene}}}\}$, is compressed into an interpretable sound event bottleneck $E$ to predict the acoustic scene $S$.}
    
    \label{fig:our_model_info_extended}
\end{figure}

Under this formal mapping, the event-related loss $\mathcal{L}_{\text{event}}$ governs the mutual information $I(X; E)$ between the raw acoustic features and the learned event concepts, whereas the scene-related loss $\mathcal{L}_{\text{scene}}$ optimizes the information $I(E; S)$ transmitted from the event bottleneck to the final scene predictions. Fundamentally, in our proposed hierarchical topology, the scene classification layer receives the predicted event probability distribution $E$ as its input, without direct access to the raw acoustic features $X$. This architecture physically forces the information flow through the sequence $X \to E \to S$, as illustrated in Fig. \ref{fig:our_model_info_extended}.

For evaluation purposes, while the hierarchical architecture establishes an explicit physical bottleneck, the unconstrained multihead baseline outputs event $E$ and scene $S$ predictions via parallel heads branching from a shared feature representation. To enable a fair comparison, we extract the event probability matrix synthesized by the multihead model and construct its conditional event distribution $P(E \mid S)$ post-hoc. This allows us to examine the implicit information bottleneck formed by the multihead joint space, evaluating whether its learned event representations naturally compress and structure scene-relevant information under the same information-theoretic metrics.

\subsection{Event-Scene Representation Framework}

A central question in our evaluation is how acoustic scene information is represented and structured through the event-level predictions of multitask architectures. If acoustic events act as an interpretable intermediate representation for scene classification—aligning with the Information Bottleneck perspective—analyzing the statistical relationships between event predictions and target scenes reveals how information is retained and structured across different model topologies.

\begin{enumerate}

    \item \textbf{Even-Scene Co-occurrence Matrix.}
        Let $E$ denote the set of acoustic event classes and $S$ the set of acoustic scene classes. To evaluate the event predictions generated by a model, we compute an aggregated event--scene co-occurrence matrix $\mathbf{C} \in \mathbb{R}^{|S| \times |E|}$. For each scene $s \in S$, we sum the predicted sigmoidal event probabilities across all $N_s$ test samples belonging to ground-truth scene $s$:

        \begin{equation}
        C_{s,e} = \sum_{n \in \mathcal{D}_s} \hat{p}_{n,e}
        \end{equation}
        
        where $\mathcal{D}_s$ is the set of audio segments assigned to scene $s$, and $\hat{p}_{n,e} \in [0, 1]$ is the predicted probability of event $e$ for segment $n$.

    \item \textbf{Empirical Conditional Event Representation.}
        To evaluate information-theoretic and geometric properties, we construct the empirical conditional event distribution $\mathbf{P} \in \mathbb{R}^{|S| \times |E|}$ directly from the co-occurrence matrix. By normalizing each scene's total event activity, the conditional probability $P(e|s)$ of detecting event $e$ given scene $s$ is defined as:
        
        \begin{equation}
        \label{eq:conditional_event_prob}
        P(e|s) = \frac{C_{s,e}}{\sum_{e' \in E} C_{s,e'}}
        \end{equation}
        
        This normalized matrix $\mathbf{P}$ reflects the fundamental acoustic profile synthesized by the multitask network, capturing both omnipresent and scene-specific event distributions in a probabilistically grounded method.

    \item \textbf{TF-IDF Event Weighting for Concept Lattices.}
    While $P(e|s)$ provides a probabilistic representation of the event-scene relationships, common background sounds may appear across many scenes and make these relationships less clear. To focus on the most discriminative acoustic events for FCA, we apply TF-IDF to the co-occurrence matrix $C$.

    Here, Term Frequency $\text{TF}(s,e)$, that is equal to $P(e|s)$, represents the normalized activity of event $e$ in scene $s$, while Inverse Document Frequency $\text{IDF}(e)$ reduces the weight of events that occur across many scene classes:
    
    \begin{equation}   
    \text{IDF}(e) = \log \left( \frac{\vert{}S\vert{}}{\vert{}\{s \in S : C_{s,e} > 0\}\vert{}} \right)
    \end{equation}
    The resulting matrix $W \in \mathbb{R}^{|S| \times |E|}$ is defined as $W_{s,e} = \text{TF}(s,e) \times \text{IDF}(e)$. Thus, events that are common across scenes receive lower weights, while more scene-specific events receive higher weights. 
    
\end{enumerate}

\section{Experimental Settings}
\label{section:experimental_settings}

\subsection{Experimental Protocols and Cross-Validation}

We segmented the audio into $5$-second windows with a $1$-second step to capture general events, resulting in a temporal resolution of $1$ second. Audio was normalized and converted to mono. The development and evaluation splits were kept in their original configurations.

For training, we used the original weights of PaSST\footnote{We use \texttt{passt\_s\_swa\_p16\_128\_ap476} as a checkpoint: \url{https://github.com/kkoutini/PaSST}}, BEATs\footnote{We use \texttt{BEATs\_iter3\_plus\_AS2M\_finetuned\_on\_AS2M\_cpt1} as a checkpoint: \url{https://github.com/microsoft/unilm/tree/master/beats}}, and PANNs\footnote{We use \texttt{Cnn14\_16k\_mAP=0.438} as a checkpoint: \url{https://github.com/bakhtos/PANNs}} pre-trained on AudioSet. 
Both the replaced final layer and the added feed-forward network (FFN) were randomly initialized, and the entire architecture was fine-tuned end-to-end for $300$ epochs with early-stopping criteria. The parameters of the pre-trained backbone were updated with a learning rate of $1 \times 10^{-5}$, while the final layers were trained with a learning rate of $1 \times 10^{-4}$. We used the Adam optimizer with a weight decay of $1 \times 10^{-4}$. The batch size was fixed at $8$ for the training of PANNs and BEATs, while in the training of PaSST was $4$. The training data was randomly shuffled during training.

For each configuration, the $\beta$ parameter explained in Section \ref{section_methods} was cross-validated, resulting in a total of 66 experiments being launched. Experiments were conducted on the computing cluster of the Department of Signal Theory and Communications at Universidad Carlos III de Madrid, equipped with multiple NVIDIA GPUs.

\subsection{Decision Thresholding and Pooling Strategies}

From the soft outputs generated by the models, posterior probabilities are obtained using a softmax function for the multi-class ASC task and an element-wise sigmoid activation for the multi-label AT task. 

For AT, frame-level predictions are first aggregated to the audio level using temporal pooling operations: Max-Pooling (capturing the occurrence of an event anywhere within the clip) or Mean-Pooling (measuring average continuous activation). To convert the pooled probabilities into binary decision vectors $\hat{\mathbf{y}}_{\text{AT}} \in \{0, 1\}^K$, three distinct thresholding strategies were evaluated:

\begin{enumerate}
    \item \textbf{Static Thresholding (ST)}: A standard global baseline threshold fixed at $\theta = 0.5$ across all acoustic categories.
    
    \item \textbf{Class-Specific Nonlinear Thresholding (CT)}: To account for class imbalance and varying event durations in the dataset, class-dependent thresholds $\boldsymbol{\theta}_{\text{class}}$ are dynamically computed based on the log-transformed total duration $d_k$ of event $k$:
    \begin{equation}
        \tilde{d}_k = \frac{\log(1 + d_k) - \min_j \log(1 + d_j)}{\max_j \log(1 + d_j) - \min_j \log(1 + d_j)}
    \end{equation}
    \begin{equation}
        \theta_k = \min\left(\theta_{\min} + \tilde{d}_k (\theta_{\max} - \theta_{\min}), \, 0.5\right)
    \end{equation}
    where $\theta_{\min} = 0.05$ and $\theta_{\max} = 0.75$. This formulation lowers the decision boundary for rare or brief acoustic events while applying a saturation bound at $0.5$ to prevent over-penalizing frequent classes.
    
    \item \textbf{Scene-Based Prior Thresholding (SPT)}: To leverage environment-dependent acoustic constraints, predicted event probabilities $\mathbf{p}_{\text{AT}}$ are adjusted using empirical prior probabilities $\mathbf{P}(E_k | \hat{S}_{\text{ASC}})$, representing the duration-normalized co-occurrence of event $k$ within the predicted scene $\hat{S}_{\text{ASC}}$:
    \begin{equation}
        \mathbf{p}_{\text{corrected}} = \alpha \mathbf{p}_{\text{AT}} + (1 - \alpha) \mathbf{P}(E | \hat{S}_{\text{ASC}})
    \end{equation}
    where $\alpha = 0.5$ balances the acoustic model confidence with empirical scene priors. Binary decisions are subsequently obtained using $\mathbf{p}_{\text{corrected}} > 0.5$.
\end{enumerate}

On the other hand, for ASC, frame-level predictions are also aggregated to the audio level using temporal pooling operations: Max-Pooling, Mean-Pooling, and Majority Vote. Additionally, there was also the option of not performing any type of pooling to remain consistent with the literature \cite{Zhang2024, zhang2025scene}.

Hyperparameters were selected via a two-stage procedure. First, to reduce computational cost, candidate epochs were provisionally fixed per loss weight ($\beta$) based on total loss minimization. At these temporary epochs, the optimal structural and post-processing parameters—ASC pooling, AT pooling, and AT threshold—were determined by joint maximization of ASC accuracy and AT F1-macro, fixing these three hyperparameters for all subsequent steps. Second, performance matrices for ASC accuracy and AT F1-macro across all epochs and $\beta$ values were recalculated using the fixed configuration. The optimal $(\beta, \text{epoch})$ pair was then selected by minimizing the weighted Euclidean distance to the ideal performance, $\sqrt{0.4 \cdot (1 - \text{ASC})^2 + 0.6 \cdot (1 - \text{AT})^2}$, computed over the min-max normalized ASC accuracy and AT F1-macro scores. This is further explained in Appendix~\ref{appendix:hyperparameters}.

\subsection{Evaluation Metrics}

\subsubsection{Overall performance}

To compare our results with those in the literature \cite{zhang2025scene}, we use accuracy as the main metric for ASC's global results. Additionally, we add balanced accuracy as a complementary metric to analyze global performance across minority classes.  
Similarly, we measure AT results using the macro F1-score and the Error Rate (ER).
To represent class performance, we use precision and recall for ASC, and the F1-score and ER for AT. The hyperparameters selected for each configuration are shown in the Appendix \ref{appendix:hyperparameters}.

\subsubsection{Information-Theoretic Evaluation Metrics}

To characterize the representation space formed by the normalized event probabilities $P(e|s)$, we evaluate three complementary information-theoretic measures:

\begin{enumerate}
    \item \textbf{Conditional Event Entropy}. To quantify the dispersion or focus of event predictions within a given scene $s$, we calculate the conditional Shannon entropy:
        \begin{equation}
        H(E \mid S=s) = -\sum_{e \in E} P(e|s) \log_2 P(e|s)
        \end{equation}
        We report the mean entropy across all scenes, $\bar{H}(E|S) = \frac{1}{|S|}\sum_{s \in S} H(E \mid S=s)$. Lower entropy values indicate a more concentrated, scene-specific event representation, whereas higher values indicate a diffuse distribution over multiple event categories.
    \item \textbf{Mutual Information}. To evaluate the total shared information between predicted events and scene categories, we compute the mutual information:
        \begin{equation}
        I(E; S) = \sum_{s \in S} \sum_{e \in E} P(s,e) \log_2 \frac{P(s,e)}{P(s)P(e)}
        \end{equation}
        where $P(s,e)$ represents the empirical joint probability distribution normalized over the full co-occurrence matrix, and $P(s) = \sum_{e \in E} P(s,e)$ and $P(e) = \sum_{s \in S} P(s,e)$ are the respective empirical marginal distributions. Higher mutual information implies a stronger statistical dependence between predicted events and target scenes, indicating that the intermediate event output retains discriminative scene information.
        
    \item \textbf{Kullback--Leibler Divergence}. To evaluate structural differences in the representations learned by the Hierarchical ($P_{\mathrm{H}}$) and Multihead ($P_{\mathrm{M}}$) architectures, we measure the divergence between their respective conditional distributions:
        \begin{equation}
        D_{\mathrm{KL}}(P_{\mathrm{H}} \parallel P_{\mathrm{M}}) = \frac{1}{|S|} \sum_{s \in S} \sum_{e \in E} P_{\mathrm{H}}(e|s) \log_2 \frac{P_{\mathrm{H}}(e|s)}{P_{\mathrm{M}}(e|s)}
        \end{equation}
        A larger KL divergence indicates that the two architectural paradigms organize intermediate event predictions across scenes in fundamentally distinct ways.
\end{enumerate}

All three measures are computed independently for each $\beta$ value and backbone architecture. This analysis provides a complementary view of the multitask models beyond predictive performance, allowing us to examine how the loss-weighting parameter influences the relationship between acoustic events and scenes and whether the two architectures learn different patterns of information sharing.

\subsubsection{t-SNE and Silhouette Score}
To evaluate the spatial organization and clustering quality of the conditional event distributions $P(E \mid S)$, we project the high-dimensional scene vectors into a two-dimensional manifold using t-Distributed Stochastic Neighbor Embedding (t-SNE). To quantitatively assess cluster compactness and separability in the resulting space, we compute the Silhouette Score, defined for each sample $i$ as $S_i = (b_i - a_i) / \max(a_i, b_i)$, where $a_i$ is the mean intra-cluster distance and $b_i$ is the mean nearest-cluster distance. Higher average scores across scenes indicate well-separated and semantically cohesive representations.

\subsubsection{FCA Lattice Construction from TF-IDF}

To obtain the formal context required for FCA, the TF-IDF matrix $W$ is converted into a binary matrix by applying a threshold $\varphi=0.15$. Each entry is therefore set to one when the corresponding event-scene association exceeds the threshold, and to zero otherwise. The resulting binary relations are used to construct the concept lattices, which provide a compact representation of the semantic relationships between acoustic events and scene classes. This allows us to compare the structures learned by the different models with the ground-truth event-scene relationships.

\subsubsection{SHAP values}

Preliminary feature importance analysis using SHAP was also conducted. SHAP values are computed on a logistic-regression surrogate trained on the intermediate event probabilities to quantify their contribution to scene prediction. Detailed plots are provided in Appendix \ref{appendix:shap} for completeness.

\section{Results and Discussion}
\label{sec:results}

\subsection{Global and Class-wise Classification Performance}
\label{sec:R_performance}

Global results for ASC and AT are given in Tables \ref{tab:global_ASC} and \ref{tab:global_AT}, respectively. Additionally, normalized confusion matrices for ASC and co-occurrence matrices for AT are reported in Appendix \ref{appendix:cm}.
Note that no ASC pooling was applied in this subsection to maintain comparability with the literature \cite{Hou2024, Zhang2024}.
In terms of overall performance, there are no significant differences between the multihead and hierarchical architectures in the ASC task; PaSST achieves better results using the hierarchical architecture, while BEATs do the contrary, and PANNs maintains similar metrics. Regarding AT global performance, PANNs has a higher ER in the hierarchical architecture, while maintaining similar F1-macro results. Regarding the results of the PaSST model, it maintains similar ER but achieves a higher F1-macro score for multihead. Finally, BEATs achieves better AT results using the hierarchical architecture. 

On the other hand, class-wise results are presented in Table \ref{tab:class_ASC} for ASC and Table \ref{tab:class_AT} for AT. Regarding ASC, there is clear confusion with the "residential area" class, as models tend to predict many false positives, leading to very low precision. PaSST is the only model to perform well on this class, owing to its superior capability among the evaluated architectures.
Looking at the AT results by class, we cannot conclude that one architecture is strictly better than the other, since some sound events achieve higher scores with the hierarchical model while others perform better with the multihead design. Note that selected hyperparameters for each configuration are shown in Appendix \ref{appendix:hyperparameters}.

In short, given the large number of cross-validated hyperparameters and experiments, we argue that there is no significant difference between the two architectures in this dataset. Small performance variations mostly depend on the selected hyperparameters or the stopping epoch—factors that slightly alter the final numbers without reflecting a real architectural advantage. However, it will be necessary to examine how more complex databases perform for the ASC task, since these hierarchical architectures rely solely on sounds as the exclusive elements for scene prediction, which may result in the loss of the gist \cite{oliva2005gist}, thereby reducing performance.
Additionally, we find a limitation when using a single dataset with a fixed development/evaluation split, as it prevents us from claiming statistical significance while remaining comparable to previous work in the literature \cite{zhang2025scene}. 

\begin{table}
\caption{Global ASC performance in terms of accuracy and balanced accuracy}  \label{tab:global_ASC}
\centering
        \small
        \begin{tabular}{l l *{3}{c}}
            \toprule
            \textbf{Architecture} & \textbf{Metric} & \textbf{PaSST} & \textbf{PANNs} & \textbf{BEATs} \\
            \midrule
            \multirow{2}{*}{Multihead} 
                                & Accuracy & 0.89 & 0.94 & 0.94 \\
                                & Balanced Acc. & 0.90 & 0.95 & 0.95 \\
            \midrule
            \multirow{2}{*}{Hierarchical} 
                                & Accuracy & 0.95 & 0.95 & 0.91 \\
                                & Balanced Acc. & 0.95 & 0.96 & 0.92 \\
            \bottomrule
        \end{tabular}
\end{table}


\begin{table} 
\caption{Global AT performance in terms of F1-macro and ER} \label{tab:global_AT}
\centering
        \small
        \begin{tabular}{l l *{3}{c}}
            \toprule
            \textbf{Architecture} & \textbf{Metric} & \textbf{PaSST} & \textbf{PANNs} & \textbf{BEATs} \\
            \midrule
            \multirow{2}{*}{Multihead}
                                & F1-macro & 0.69 & 0.61 & 0.62 \\
                                & ER      & 0.27 & 0.27 & 0.35 \\
            \midrule
            \multirow{2}{*}{Hierarchical}
                                & F1-macro & 0.63 & 0.63 & 0.66 \\
                                & ER      & 0.28 & 0.38 & 0.31 \\
            \bottomrule
        \end{tabular}
\end{table}


\begin{table}
    \caption{Class-wise ASC performance in terms of precision and recall} 
    \label{tab:class_ASC}
    \centering
        \setlength{\tabcolsep}{3pt}
        \begin{tabular}{l l l *{4}{c}}
            \toprule
            \textbf{Arch.} & \textbf{Model} & \textbf{Metric} & \textbf{home} & \textbf{office} & \textbf{city} & \textbf{residential} \\
            & & &  &  & \textbf{center} & \textbf{area} \\
            \midrule
            \multirow{6}{*}{Multihead}             & \multirow{2}{*}{PaSST} & Precision & 0.99 & 1.00 & 0.80 & 0.82 \\
                                       &                        & Recall    & 1.00 & 0.98 & 0.85 & 0.77 \\
                                \cmidrule{2-7}
                                                   & \multirow{2}{*}{PANNs} & Precision & 0.99 & 1.00 & 0.96 & 0.85 \\
                                       &                        & Recall    & 0.99 & 0.98 & 0.84 & 0.97 \\
                                \cmidrule{2-7}
                                                   & \multirow{2}{*}{BEATs} & Precision & 0.99 & 1.00 & 0.99 & 0.84 \\
                                       &                        & Recall    & 1.00 & 0.98 & 0.83 & 1.00 \\
            \midrule
            \multirow{6}{*}{Hierarchical}          & \multirow{2}{*}{PaSST} & Precision & 0.99 & 1.00 & 0.98 & 0.86 \\
                                       &                        & Recall    & 1.00 & 0.98 & 0.85 & 0.98 \\
                                \cmidrule{2-7}
                                                   & \multirow{2}{*}{PANNs} & Precision & 1.00 & 1.00 & 0.93 & 0.90 \\
                                       &                        & Recall    & 0.99 & 0.99 & 0.90 & 0.93 \\
                                \cmidrule{2-7}
                                                   & \multirow{2}{*}{BEATs} & Precision & 1.00 & 0.99 & 0.81 & 0.89 \\
                                       &                        & Recall    & 0.99 & 1.00 & 0.91 & 0.78 \\
            \bottomrule
        \end{tabular}
\end{table}

\begin{table*}
\caption{Class-wise AT performance in terms of F1-score and ER} \label{tab:class_AT}
\centering
        \setlength{\tabcolsep}{3pt}
        \begin{tabular}{l l l *{9}{c}}
            \toprule
            \textbf{Architecture} & \textbf{Model} & \textbf{Metric} & \textbf{banging} & \textbf{impact} & \textbf{rustling} & \textbf{snapping} & \textbf{squeaking} & \textbf{bird} & \textbf{brakes} & \textbf{breathing} & \textbf{car} \\
            & & & \textbf{(obj.)} & \textbf{(obj.)} & \textbf{(obj.)} & \textbf{(obj.)} & \textbf{(obj.)} & \textbf{singing} & \textbf{squeaking} &  &  \\
            \midrule
            \multirow{6}{*}{Multihead}
                                & \multirow{2}{*}{PaSST} & F1-score & 0.75 & 1.00 & 0.77 & 0.91 & 0.00 & 0.63 & 0.73 & 0.33 & 0.83 \\
                                &                        & ER      & 0.25 & 0.00 & 0.25 & 0.17 & 1.00 & 0.51 & 0.28 & 0.33 & 0.27 \\
                                \cmidrule{2-12}
                                & \multirow{2}{*}{PANNs} & F1-score & 0.40 & 1.00 & 0.52 & 0.57 & 0.00 & 0.73 & 0.54 & 0.22 & 0.82 \\
                                &                        & ER      & 0.00 & 0.00 & 0.25 & 0.00 & 1.00 & 0.32 & 0.23 & 0.78 & 0.29 \\
                                \cmidrule{2-12}
                                & \multirow{2}{*}{BEATs} & F1-score & 0.40 & 1.00 & 0.80 & 0.75 & 0.00 & 0.70 & 0.70 & 0.17 & 0.82 \\
                                &                        & ER      & 0.00 & 0.00 & 0.30 & 0.00 & 1.00 & 0.33 & 0.21 & 0.87 & 0.29 \\
            \midrule
            \multirow{6}{*}{Hierarchical}
                                & \multirow{2}{*}{PaSST} & F1-score & 0.67 & 1.00 & 0.71 & 0.89 & 0.00 & 0.67 & 0.67 & 0.15 & 0.83 \\
                                &                        & ER      & 0.00 & 0.00 & 0.23 & 0.00 & 1.00 & 0.45 & 0.13 & 0.75 & 0.28 \\
                                \cmidrule{2-12}
                                & \multirow{2}{*}{PANNs} & F1-score & 0.53 & 1.00 & 0.77 & 1.00 & 0.33 & 0.65 & 0.54 & 0.15 & 0.84 \\
                                &                        & ER      & 0.64 & 0.00 & 0.09 & 0.00 & 0.00 & 0.47 & 0.62 & 0.75 & 0.28 \\
                                \cmidrule{2-12}
                                & \multirow{2}{*}{BEATs} & F1-score & 0.86 & 0.98 & 0.74 & 1.00 & 0.00 & 0.70 & 0.62 & 0.13 & 0.82 \\
                                &                        & ER      & 0.00 & 0.00 & 0.17 & 0.00 & 1.00 & 0.41 & 0.33 & 0.83 & 0.27 \\
            \midrule \addlinespace[2pt] \midrule
            \textbf{Architecture} & \textbf{Model} & \textbf{Metric} & \textbf{children} & \textbf{cupboard} & \textbf{cutlery} & \textbf{dishes} & \textbf{drawer} & \textbf{fan} & \textbf{glass} & \textbf{keyb.} & \textbf{large.} \\
            & & &  &  &  &  &  &  & \textbf{jingling} & \textbf{typ.} & \textbf{vehic.} \\
            \midrule
            \multirow{6}{*}{Multihead}
                                & \multirow{2}{*}{PaSST} & F1-score & 0.18 & 0.89 & 0.57 & 0.89 & 0.89 & 1.00 & 0.50 & 0.67 & 0.39 \\
                                &                        & ER      & 0.00 & 0.00 & 0.50 & 0.20 & 0.00 & 0.00 & 0.00 & 0.50 & 0.71 \\
                                \cmidrule{2-12}
                                & \multirow{2}{*}{PANNs} & F1-score & 0.00 & 0.57 & 0.67 & 0.89 & 0.57 & 0.98 & 0.80 & 0.67 & 0.39 \\
                                &                        & ER      & 1.00 & 0.00 & 0.33 & 0.20 & 0.00 & 0.04 & 0.00 & 0.48 & 0.62 \\
                                \cmidrule{2-12}
                                & \multirow{2}{*}{BEATs} & F1-score & 0.00 & 0.57 & 0.75 & 0.57 & 0.75 & 0.98 & 0.00 & 0.72 & 0.26 \\
                                &                        & ER      & 1.00 & 0.00 & 0.40 & 0.33 & 0.00 & 0.04 & 1.00 & 0.43 & 0.78 \\
            \midrule
            \multirow{6}{*}{Hierarchical}
                                & \multirow{2}{*}{PaSST} & F1-score & 0.18 & 0.57 & 0.40 & 0.67 & 0.75 & 0.98 & 0.00 & 0.65 & 0.35 \\
                                &                        & ER      & 0.00 & 0.00 & 0.50 & 0.00 & 0.00 & 0.04 & 1.00 & 0.52 & 0.74 \\
                                \cmidrule{2-12}
                                & \multirow{2}{*}{PANNs} & F1-score & 0.61 & 1.00 & 0.75 & 0.26 & 1.00 & 0.96 & 0.40 & 0.62 & 0.33 \\
                                &                        & ER      & 0.46 & 0.00 & 0.40 & 0.85 & 0.00 & 0.07 & 0.50 & 0.55 & 0.80 \\
                                \cmidrule{2-12}
                                & \multirow{2}{*}{BEATs} & F1-score & 0.18 & 0.89 & 0.75 & 0.89 & 0.57 & 0.98 & 0.40 & 0.69 & 0.37 \\
                                &                        & ER      & 0.00 & 0.00 & 0.40 & 0.20 & 0.00 & 0.04 & 0.50 & 0.45 & 0.73 \\
            \midrule \addlinespace[2pt] \midrule
            \textbf{Architecture} & \textbf{Model} & \textbf{Metric} & \textbf{mouse} & \textbf{mouse} & \textbf{peop.} & \textbf{peop.} & \textbf{wash.} & \textbf{water} & \textbf{wind} \\
            & & & \textbf{click.} & \textbf{wheel.} & \textbf{talk.} & \textbf{walk.} & \textbf{dish.} & \textbf{tap} & \textbf{bl.} \\
            \midrule
            \multirow{6}{*}{Multihead}
                                & \multirow{2}{*}{PaSST} & F1-score & 0.98 & 0.47 & 0.63 & 0.52 & 0.89 & 1.00 & 0.75 \\
                                &                        & ER      & 0.00 & 0.33 & 0.45 & 0.39 & 0.20 & 0.00 & 0.40 \\
                                \cmidrule{2-12}
                                & \multirow{2}{*}{PANNs} & F1-score & 0.96 & 0.27 & 0.48 & 0.45 & 0.89 & 1.00 & 0.80 \\
                                &                        & ER      & 0.08 & 0.50 & 0.33 & 0.23 & 0.20 & 0.00 & 0.00 \\
                                \cmidrule{2-12}
                                & \multirow{2}{*}{BEATs} & F1-score & 0.96 & 0.62 & 0.68 & 0.62 & 0.89 & 1.00 & 0.67 \\
                                &                        & ER      & 0.00 & 0.52 & 0.41 & 0.33 & 0.20 & 0.00 & 0.33 \\
            \midrule
            \multirow{6}{*}{Hierarchical}
                                & \multirow{2}{*}{PaSST} & F1-score & 0.96 & 0.44 & 0.59 & 0.62 & 0.89 & 1.00 & 1.00 \\
                                &                        & ER      & 0.08 & 0.43 & 0.52 & 0.24 & 0.20 & 0.00 & 0.00 \\
                                \cmidrule{2-12}
                                & \multirow{2}{*}{PANNs} & F1-score & 0.94 & 0.54 & 0.38 & 0.47 & 0.73 & 1.00 & 0.00 \\
                                &                        & ER      & 0.11 & 0.53 & 0.50 & 0.55 & 0.43 & 0.00 & 1.00 \\
                                \cmidrule{2-12}
                                & \multirow{2}{*}{BEATs} & F1-score & 0.98 & 0.33 & 0.61 & 0.56 & 0.89 & 1.00 & 0.57 \\
                                &                        & ER      & 0.04 & 0.57 & 0.53 & 0.49 & 0.20 & 0.00 & 0.50 \\
            
            \bottomrule
        \end{tabular}
\end{table*}

\subsection{Training Dynamics across Loss Weighting ($\beta$)}

\begin{figure}
    \centering  
    \includegraphics[width=1\linewidth]{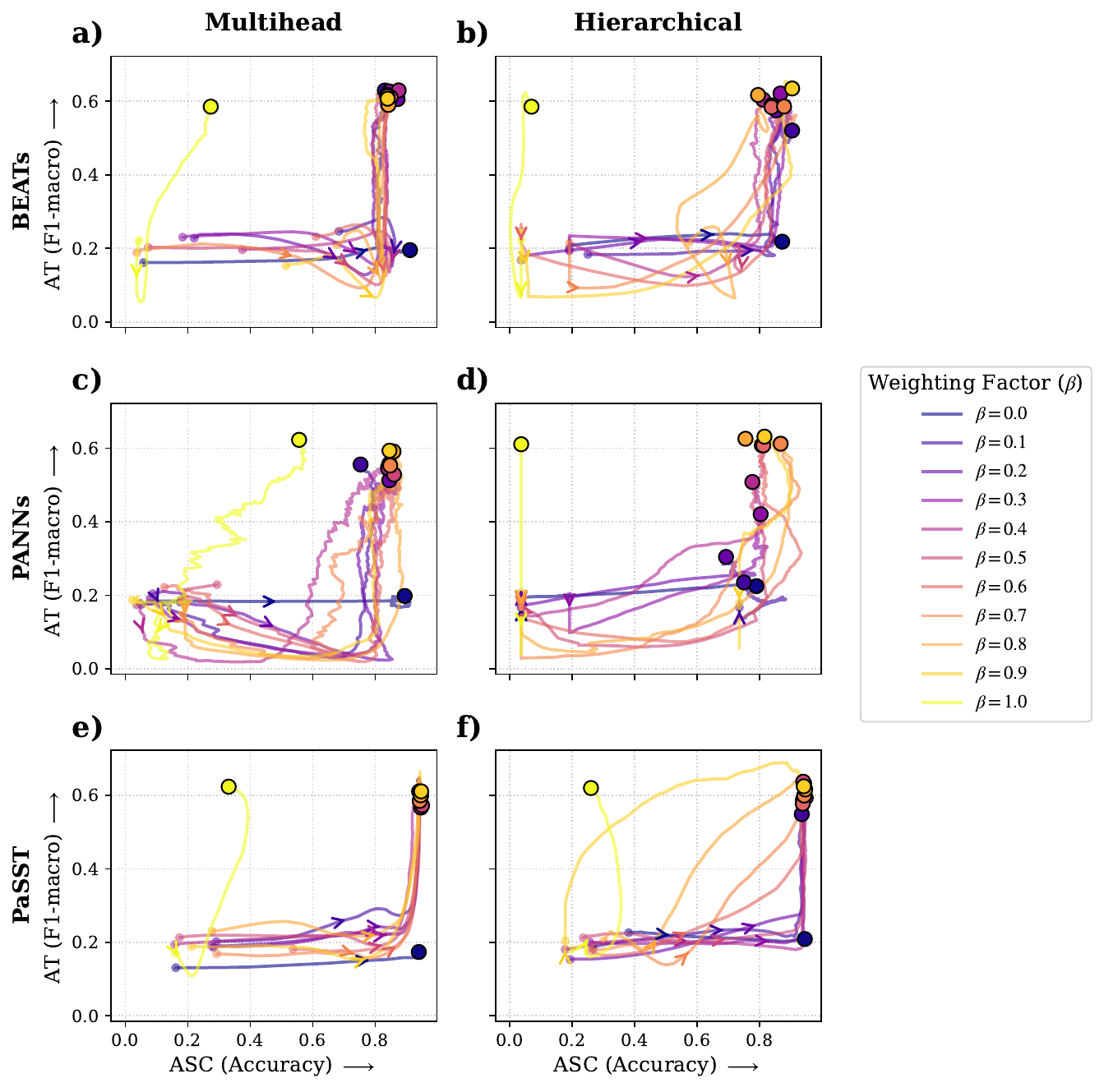}
    \caption{Test performance trajectories of the multitask models across different task-loss weights $\beta$, represented by color, in the ASC accuracy -- AT F1-macro space. The left column \textbf{(a,c,e)} shows the multihead architecture, while the right column \textbf{(b,d,f)} shows the Hierarchical architecture. From top to bottom, rows correspond to BEATs, PANNs, and PaSST, respectively. Each trajectory connects the test performance obtained for the different $\beta$ values, with a $90\%$ smoothing applied for visualization.} \label{fig:beta}
    \label{fig:placeholder}
\end{figure}

Fig. \ref{fig:beta} shows the evaluation performance over training epochs in terms of accuracy for ASC and macro F1-score for AT. Each trajectory represents a specific $\beta$ value, illustrating how the weighting factor influences training dynamics.

These trajectories reveal interesting patterns from an information-compression perspective. For the multihead architecture, the model initially prioritizes the simpler task, ASC. Once a baseline representation for ASC is established, it begins learning the more complex task, AT. Visually, this sequential optimization causes the performance paths to shift rightward first along the ASC axis before moving upward along the AT axis.

A key finding of this study is that this decoupled learning behavior does not occur in the hierarchical architecture, particularly for higher $\beta$ values ($0.6 \le \beta \le 0.9$) that favor the AT task. In these cases, performance on both tasks increases simultaneously, indicating that learning AT features directly drives and optimizes the underlying representation needed for ASC.
This joint convergence suggests that the hierarchical bottleneck forces the model to compress event-level details into a shared representation that immediately benefits scene inference, rather than learning two isolated tasks as observed in the multihead setup.

Additionally, Fig.~\ref{fig:beta} further supports our previous observation regarding the lack of a clear performance gap between multihead and hierarchical architectures. The endpoints of the trajectories—representing peak performance across both metrics—reach very similar regions in the task-space. Minor advantages in either ASC accuracy or AT F1-score are heavily dependent on early stopping criteria or specific hyperparameter setups rather than an intrinsic architectural superiority.

\subsection{Information Bottleneck and Representation Profiling}

\begin{figure*}
    \centering
    \includegraphics[width=1\textwidth]{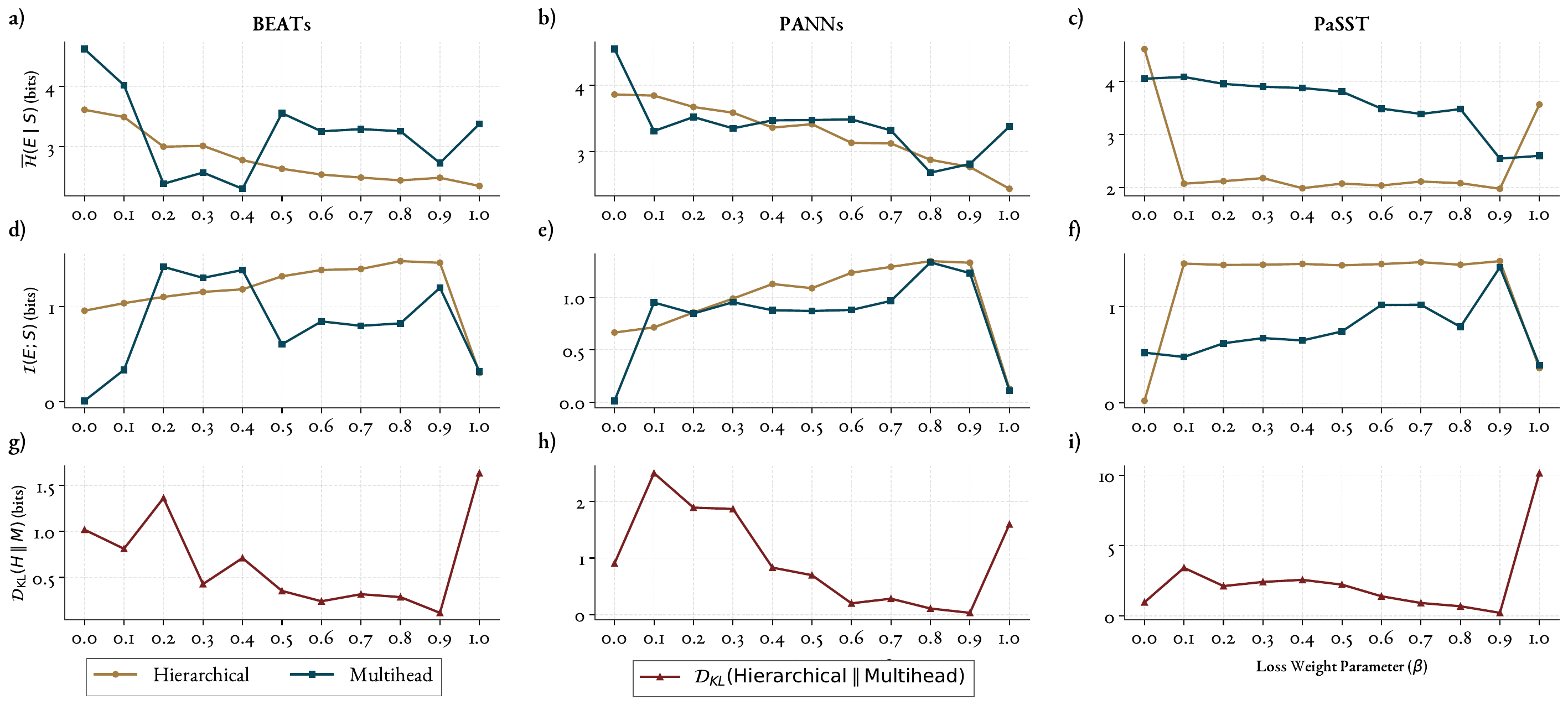}
    \caption{Information-theoretic comparison of the learned event representations across BEATs, PANNs, and PaSST. Rows show mean conditional event entropy $\bar{H}(E\mid S)$, mutual information $I(E;S)$, and Kullback-Leibler divergence $D_{\mathrm{KL}}(P_H|P_M)$, respectively. Colors distinguish the hierarchical and multihead architectures.}
    \label{fig:information}
\end{figure*}

To evaluate how intermediate event representations organize and transmit scene-discriminative information, Fig. \ref{fig:information} illustrates the mean conditional entropy $\bar{H}(E \mid S)$, mutual information $I(E; S)$, and Kullback--Leibler divergence $D_{\mathrm{KL}}(P_{\mathrm{H}} \parallel P_{\mathrm{M}})$ across all three model backbones (PaSST, PANNs, and BEATs) as a function of the multitask loss weight $\beta$.

Across almost all models and hyperparameter configurations ($\beta$), the proposed hierarchical architecture consistently yields a lower conditional mean entropy $\bar{H}(E \mid S)$ compared to the multihead baseline. In accordance with the information bottleneck principle, this lower entropy indicates that hierarchical conditioning effectively acts as a structural regularizer, suppressing diffuse background event activations and focusing the probability mass onto a compact, scene-specific subset of acoustic triggers.

Additionally, this spatial compression does not degrade semantic fidelity. As shown in Fig. \ref{fig:information}, the hierarchical architecture achieves systematically higher mutual information $I(E; S)$ across most $\beta$ values. This confirms that while the hierarchical mechanism prunes irrelevant event noise, it preserves the statistical dependence between the intermediate event predictions and the target scene categories. Consequently, the intermediate event layer serves as a highly discriminative acoustic bottleneck.

Finally, the Kullback-Leibler divergence $D_{\mathrm{KL}}(P_{\mathrm{H}} \parallel P_{\mathrm{M}})$ evaluates the structural distance between the final converged event distributions of both architectures. Across all models, $D_{\mathrm{KL}}$ reaches a minimum at $\beta = 0.9$. While the training trajectories of the hierarchical and multihead models differ significantly throughout optimization (shown in Fig. \ref{fig:beta}), their final converged representations become remarkably similar under this configuration. At $\beta = 0.9$, event loss dominates the total loss. This forces both architectures to prioritize fitting raw event labels, minimizing the structural effect of the hierarchical head and yielding almost identical final event distributions. However, for $\beta < 0.9$, the hierarchical constraint successfully shapes the representation space, keeping it structurally distinct from the multihead baseline.

\subsection{Manifold Geometry and Cluster Separability}

\begin{figure}
    \centering
    \includegraphics[width=1\linewidth]{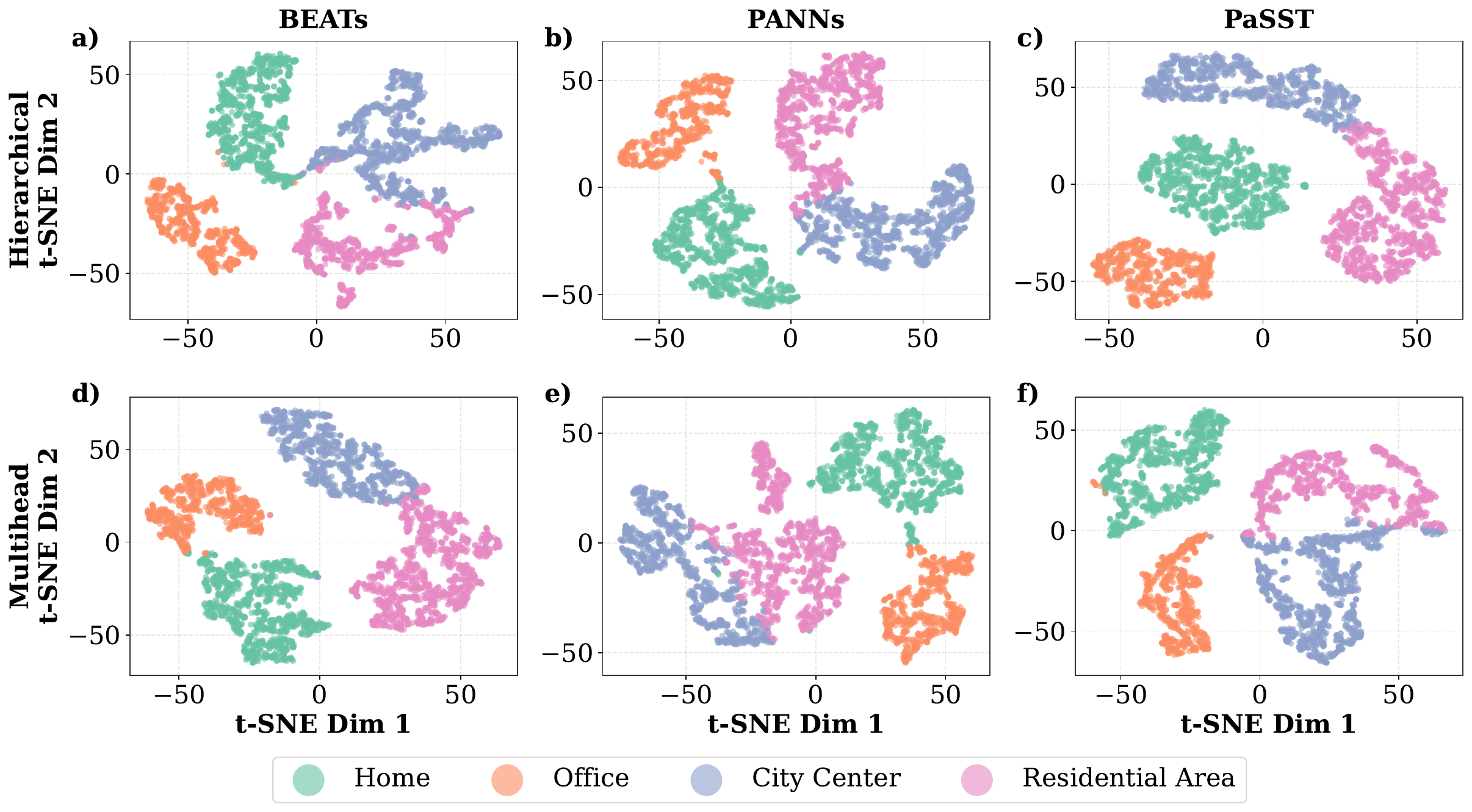}
    \caption{Two-dimensional t-SNE projections of the predicted event probability distributions $P(E\mid S)$, where each point represents an audio segment, and colors indicate the ground-truth acoustic scene. The projections illustrate the spatial organization and cluster separability of the learned event representations across the different architectures and pretrained audio backbones.}
    \label{fig:tsne}
\end{figure}

To complement the information-theoretic analysis, we examine the spatial organization of the learned event space by treating the predicted event probabilities $P(E \mid S)$ as feature vectors. Fig. \ref{fig:tsne} presents a two-dimensional t-SNE projection, where each point corresponds to an audio segment colored by its ground-truth acoustic scene. 

Although t-SNE is a stochastic dimensionality reduction technique, the overall topological structure remains highly consistent. Across both multihead and hierarchical architectures, the representations form well-defined, spatially separated clusters corresponding to distinct acoustic environments. Small overlaps occur mostly between "city center" and "residential area", reflecting their shared outdoor background events (such as "car" or "children").

To objectively quantify cluster compactness and boundary separation independent of t-SNE stochasticity, Table \ref{tab:silohuette} presents the Silhouette Scores across all three model backbones. For PaSST and PANNs, the hierarchical architecture achieves noticeably higher Silhouette Scores compared to the multihead baseline, confirming that hierarchical conditioning creates more segregated scene clusters in latent space. 

In contrast, BEATs shows that the multihead baseline yields a higher score than its hierarchical architecture. This aligns with our previous observations in Fig. \ref{fig:information}, where for BEATs evaluated under its optimal configuration ($\beta = 0.4$, shown in Table \ref{tab:hyperparameters}), the multihead model achieves a lower conditional entropy than the hierarchical variant at higher weights ($\beta = 0.9$, shown in Table \ref{tab:hyperparameters}). Nevertheless, taking both information-theoretic metrics and visual manifold geometry into account, the hierarchical paradigm consistently demonstrates superior or highly competitive representation structure across model backbones.

\begin{table}  
\caption{\textbf{Silhouette Scores} of the learned event representations across the three pretrained audio backbones.} \label{tab:silohuette}
\centering
        \begin{tabular}{l *{3}{c}}
            \toprule
            \textbf{Architecture} & \textbf{PaSST} & \textbf{PANNs} & \textbf{BEATs} \\
            \midrule
            {Multihead} & 0.451 & 0.426 & 0.529 \\              
            {Hierarchical} & 0.534 & 0.512& 0.478 \\   
            \bottomrule
        \end{tabular}

\end{table}  

\subsection{Semantic Concept Preservation via Formal Concept Analysis}

While information-theoretic and geometric metrics summarize global representation quality, they do not explicitly reveal which semantic associations are learned or corrupted by the model. To bridge this interpretability gap, we construct FCA concept lattices from the TF-IDF-weighted event predictions and compare them directly against the ground-truth context lattice (Fig. \ref{fig:fca_dataset} \textbf{(f)}). By minimizing background noise via TF-IDF, FCA isolates the structural hierarchy of highly discriminative acoustic concepts.

\begin{figure*}
    \centering
    \includegraphics[width=1\textwidth]{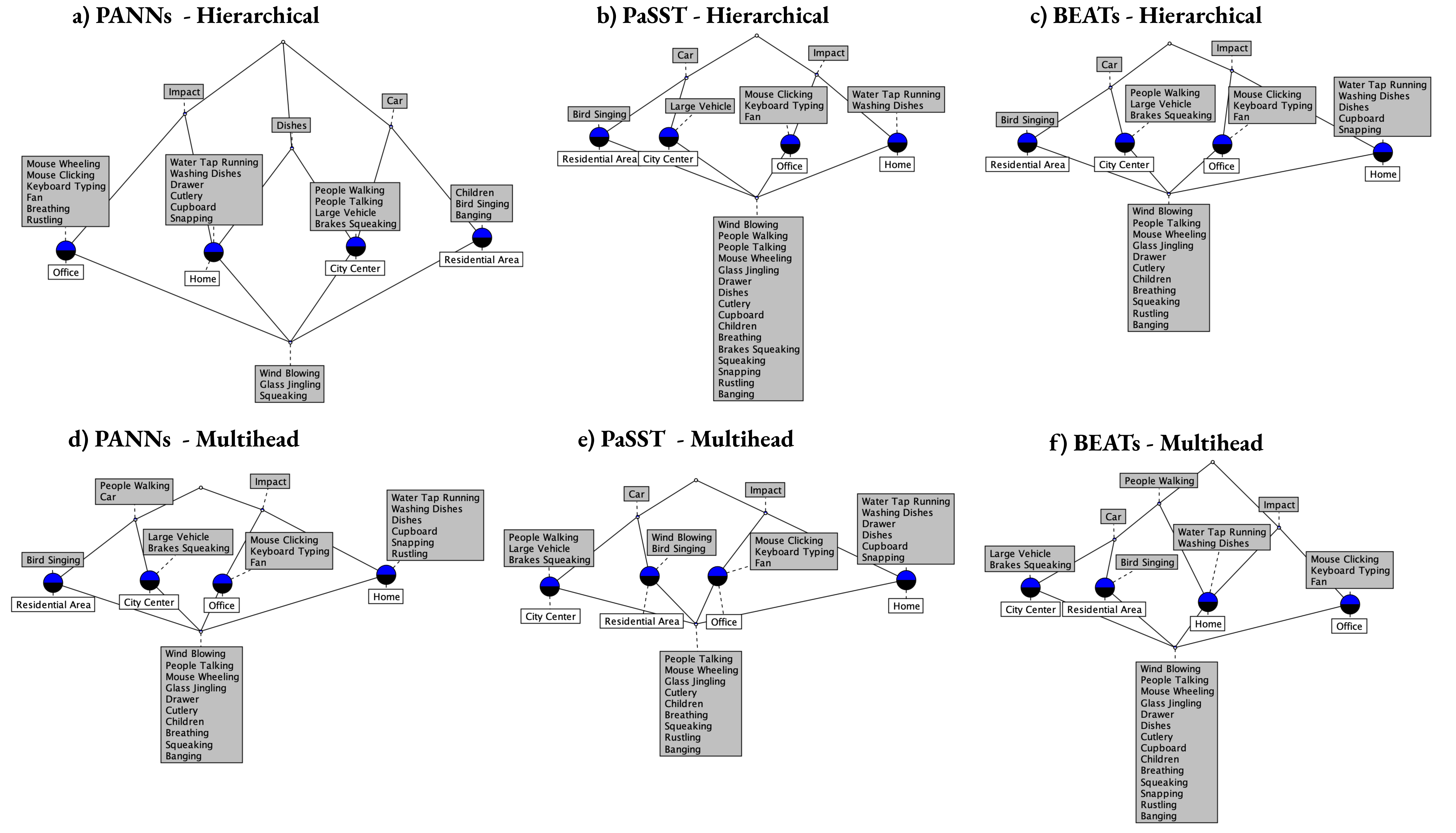}
    \caption{FCA lattices constructed from the TF-IDF-weighted event-scene relationship matrices predicted by the models, using a threshold of $\varphi=0.15$. The top row \textbf{(a-c)} corresponds to the hierarchical architecture, while the bottom row \textbf{(d-f)} corresponds to the multihead architecture. From left to right, columns correspond to PANNs, PaSST, and BEATs, respectively.}
    \label{fig:fca_lattices}
\end{figure*}

Fig. \ref{fig:fca_lattices} shows the lattices obtained from the predictions of the six models on the evaluation set, which can be compared with the ground-truth TF-IDF lattice in Fig. \ref{fig:fca_lattices} \textbf{(f)}. All predicted lattices are sparser than the ground truth: they contain 8 or 9 concepts and recover between 11 and 22 of the 28 event–scene incidences of the reference context, while introducing very few spurious associations (at most three per model). This indicates that the models behave conservatively, retaining only the associations for which they provide the strongest and most specific evidence. Nevertheless, the backbone of the ground-truth structure is preserved in all cases: "car" remains the concept shared by "city center" and "residential area", "impact" remains the concept shared by "home" and "office", and each scene keeps a set of characteristic exclusive events, such as "large vehicle" for "city center", "bird singing" for "residential area", "water tap running" and "washing dishes" for "home", and "fan", "keyboard typing", and "mouse clicking" for "office".

The main structural difference between both architectures concerns "people walking", the most widespread and therefore least discriminative event in the ground truth, where it links "home" with both outdoor scenes. In the hierarchical models, this event is never shared across scenes: it is either assigned exclusively to "city center" (PANNs and BEATs) or discarded (PaSST). In contrast, two of the three multihead models retain it as a shared attribute, either between the two outdoor scenes (PANNs) or across "home", "city center", and "residential area" (BEATs), thus reproducing the crossing link between the indoor and outdoor branches. As a result, the hierarchical lattices of PaSST and BEATs reduce to two clearly separated branches, indoor and outdoor, that only meet at the top concept. This behavior is consistent with the lower conditional entropy $H(E|S)$ observed for the hierarchical models: by concentrating the predicted event distribution of each scene on fewer and more specific events, hierarchical conditioning suppresses frequent associations but carries little information about the scene.

The amount of detail retained, however, depends on the backbone. The hierarchical PaSST model yields the most compact lattice, recovering only 11 incidences, whereas the hierarchical PANNs model yields the richest one, with 22 correct incidences and the highest agreement with the ground truth among all models. It is also the only model that recovers "children" and "banging" in "residential area" and "rustling" in "office", the latter being one of the ambiguous events identified in Section III-B. In exchange, it introduces a spurious association between "dishes" and "city center", which creates a link between "home" and "city center" absent from the ground truth. It also assigns "people talking" to "city center" and "breathing" to "office", associations that appear in the development set (Fig. \ref{fig:fca_dataset} \textbf{(c)}) but not in the evaluation set. Consequently, at the level of individual incidences, neither architecture is uniformly superior: the hierarchical models achieve higher agreement for PANNs and BEATs, while the multihead model does so for PaSST.

Finally, some ground-truth associations are missed by all models, notably "glass jingling" in "home", "mouse wheeling" in "office", and "people talking" in "residential area". "Residential area" is the scene with the largest loss, retaining at most four of its seven associated events. These correspond mostly to low-prevalence or acoustically ambiguous events, whose predicted probabilities rarely exceed $\varphi$ after TF-IDF weighting. Likewise, no model associates "bird singing" with "city center", an association that in the ground truth arises only from the normalization of a weak event (Section \ref{sec:dataset:FCA}). Overall, these results support that hierarchical conditioning preserves the key event–scene relationships of the ground truth while discarding less informative shared associations, although the level of detail retained strongly depends on the backbone.




\section{Limitations and Future Work} \label{sec:limitations_future_work}

While our systematic benchmark provides critical insights into the optimization dynamics and latent geometry of multitask acoustic models, several limitations remain to be addressed in subsequent research:

\begin{enumerate}
    \item \textbf{Dataset Lack and Domain Generalization}. High-quality datasets offering simultaneous, aligned annotations for both broad acoustic scenes and fine-grained audio events are exceptionally lacking in the audio community. Consequently, our findings are based on a single benchmark dataset. Evaluating whether these representation properties hold in larger-scale, highly unconstrained, or open-set acoustic environments remains an important next step.

    \item \textbf{Potential Loss of Acoustic Gist}. By routing scene classification exclusively through the intermediate event probability matrix $P(E \mid S)$, the hierarchical architecture enforces a strict information bottleneck. If the pre-defined event vocabulary fails to capture fine background textures—the global acoustic gist of an environment—there is a risk of error propagation from missed event triggers to final scene predictions.

    \item \textbf{Computational Overhead of Hyperparameter Sweeping}. Exhaustively exploring the joint trade-off space across multiple backbone architectures required significant compute resources (66 full training runs). Developing dynamic loss-weighting algorithms that automatically balance task learning without requiring offline grid searches represents a key optimization challenge.
    
\end{enumerate}

To build upon these findings, future work will focus on three main directions. First, we plan to extend this interpretability framework by analyzing the relationship between ASC and SED, instead of AT. Second, we aim to investigate hybrid architectures that combine the structured intermediate event bottleneck $P(E \mid S)$ with a residual global feature bypass to preserve ambient acoustic context. Finally, we will explore self-supervised pre-training strategies to expand the intermediate event vocabulary, reducing the risk of domain-specific information bottlenecks.

\section{Conclusion} \label{sec:conclusion}

In this paper, we addressed the limitations of standard unconstrained multihead architectures for joint ASC and AT by proposing a hierarchical (sequential) conditioning framework. By conditioning scene predictions directly on intermediate, normalized event probability distributions $P(E \mid S)$, our approach aligns model structure with the intrinsic hierarchy of physical acoustic environments. 

Through an extensive empirical evaluation encompassing 66 full training runs across three state-of-the-art backbones (PaSST, PANNs, and BEATs), we demonstrated that hierarchical conditioning achieves competitive overall classification metrics while fundamentally stabilizing multitask optimization. Crucially, optimization dynamics reveal that whereas multihead architectures master the simpler ASC task early before optimizing AT, the Hierarchical approach fosters synchronized dual-task learning across balanced loss weights ($\beta \in [0.6, 0.9]$). 

Furthermore, our multi-perspective interpretability framework—combining information-theoretic bottleneck metrics, manifold geometry analysis, and FCA on TF-IDF event activations—shows that the hierarchical paradigm produces a more structured intermediate representation space. Across model architectures, hierarchical conditioning consistently yields lower conditional entropy $H(E \mid S)$, higher mutual information $I(E; S)$, and tighter latent cluster separability (higher Silhouette scores). The FCA lattices further show that hierarchical models preserve key event-scene relationships from the ground truth while reducing less informative associations, providing an interpretable view of the semantic structure learned by the models.

\section{Acknowledgements}

\clearpage

\bibliographystyle{IEEEtran}
\bibliography{my_bib}

\vfill

\clearpage

\appendix

\subsection{Hyperparameters} \label{appendix:hyperparameters}

Hyperparameter configurations for each model and architecture were selected via a two-stage optimization procedure:

\begin{enumerate}
    \item \textbf{Structural and Post-Processing Parameter Selection:} To significantly reduce computational cost during grid search, candidate epochs were provisionally fixed for each multitask loss weight ($\beta$) by minimizing the total validation loss. At these tentative epochs, a cross-validation grid search was conducted over structural and post-processing settings—specifically ASC pooling, AT pooling, and AT thresholds. The optimal trio of parameters was selected by jointly maximizing ASC accuracy and AT F1-macro, and then maintained fixed for all subsequent evaluations.
    
    \item \textbf{Joint Epoch and Loss Weight Selection:} Using the fixed configuration from the first stage, full performance matrices for ASC accuracy and AT F1-macro were recalculated across all training epochs and $\beta$ values. Both metrics were min-max normalized across the entire grid:
    \begin{equation}
    M_{\text{norm}} = \frac{M - M_{\min}}{M_{\max} - M_{\min}}, \quad M \in \{\text{ASC}, \text{AT}\}
    \end{equation}
    
    The optimal joint training epoch ($e^*$) and trade-off weight ($\beta^*$) were identified by finding the configuration that minimizes the weighted Euclidean distance to ideal joint performance:
    \begin{equation}
    \begin{aligned}
    (e^*, \beta^*) = \arg\min_{e, \beta} \Big[ & 0.4 \left(1 - \text{ASC}_{e,\beta}^{\text{norm}}\right)^2 \\
    &+ 0.6 \left(1 - \text{AT}_{e,\beta}^{\text{norm}}\right)^2 \Big]^{1/2}
    \end{aligned}
    \end{equation}
    where a $0.60$ weight is assigned to the AT macro-$F_1$ score and $0.40$ to the ASC accuracy, reflecting the multi-label complexity of the event detection task.
\end{enumerate}

Table \ref{tab:hyperparameters} summarizes the selected hyperparameter configurations across all evaluated models and architectural variants.

\begin{table}[h]
    \caption{Optimal hyperparameter configurations selected per backbone model and architectural approach. Threshold strategies are abbreviated as \textbf{CT} (Class-specific Thresholding) and \textbf{ST} (Static Thresholding).}
    \label{tab:hyperparameters}
    \centering
    \small
    \begin{tabular}{l l *{5}{c}}
        \toprule
        \textbf{Arch.} & \textbf{Model} & \textbf{Epoch} & $\beta$  & \textbf{ASC}  & \textbf{AT}  & \textbf{AT}\\
         & & & & \textbf{Pool.} & \textbf{Pool.}  & \textbf{Th.} \\
        \midrule
        \multirow{3}{*}{Multihead} 
        & PaSST & 108 & 0.9 & None & max & CT \\
        & PANNs & 296 & 0.8 & max$^*$  & max & ST \\
        & BEATs & 228 & 0.4 & max$^*$  & max & ST \\
        \midrule
        \multirow{3}{*}{Hierarchical} 
        & PaSST & 227 & 0.9 & None & max & CT \\
        & PANNs & 187 & 0.7 & max$^*$  & max & ST \\
        & BEATs & 130 & 0.9 & max$^*$  & max & ST \\
        \bottomrule
    \end{tabular}

    \footnotesize $^*$Note that numerical results reported in Section \ref{sec:R_performance} were calculated using no-pooling strategy for ASC (i.e., "None").

\end{table}

\subsection{Feature Importance via SHAP Values} \label{appendix:shap}

\begin{figure}[h]
    \centering
    \includegraphics[width=1\linewidth]{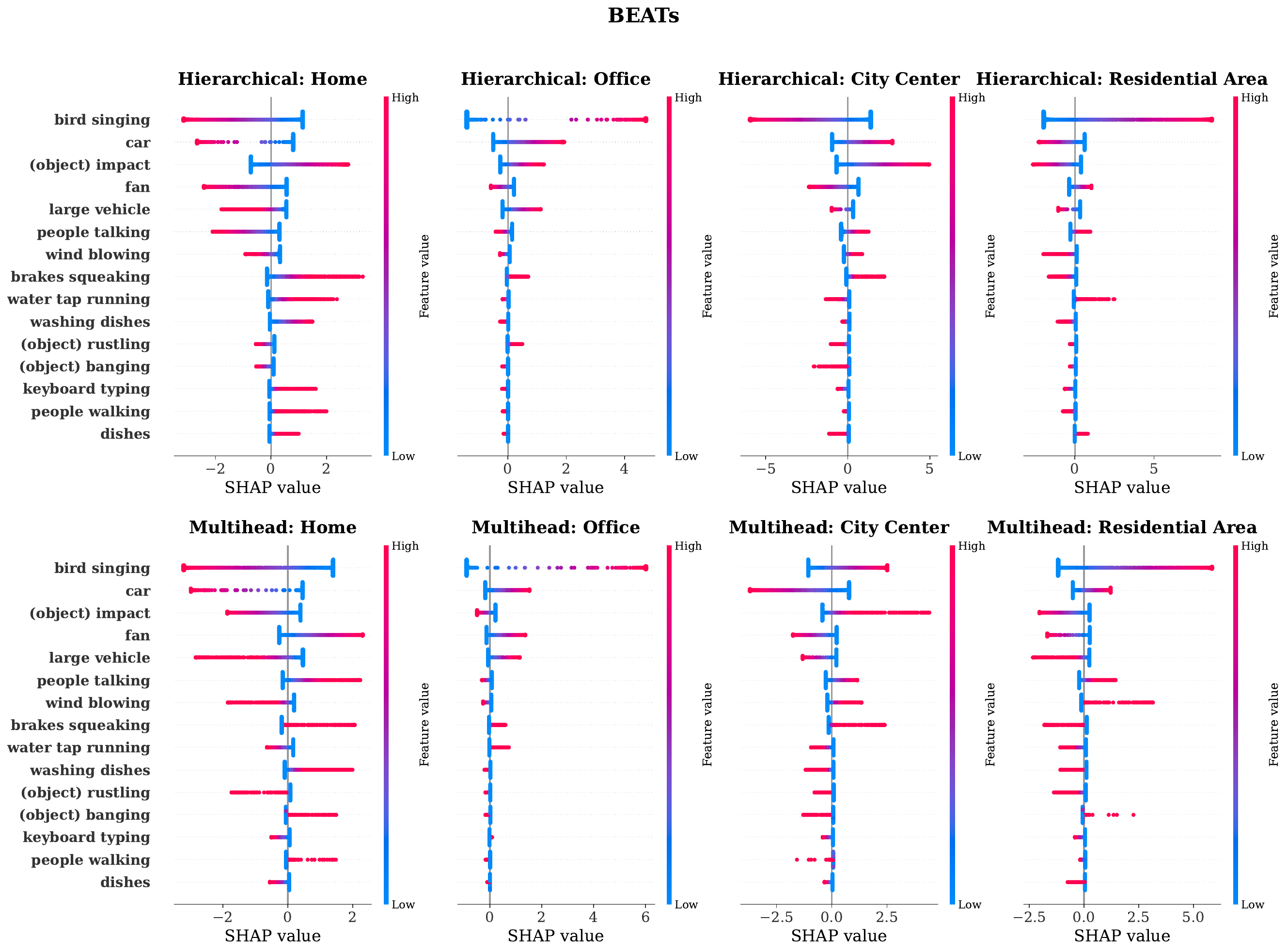}
    \caption{SHAP value summary plot comparing feature importance for BEATs under Multihead and Hierarchical configurations.}
    \label{fig:shap_beats}
\end{figure}

\begin{figure}[h]
    \centering
    \includegraphics[width=1\linewidth]{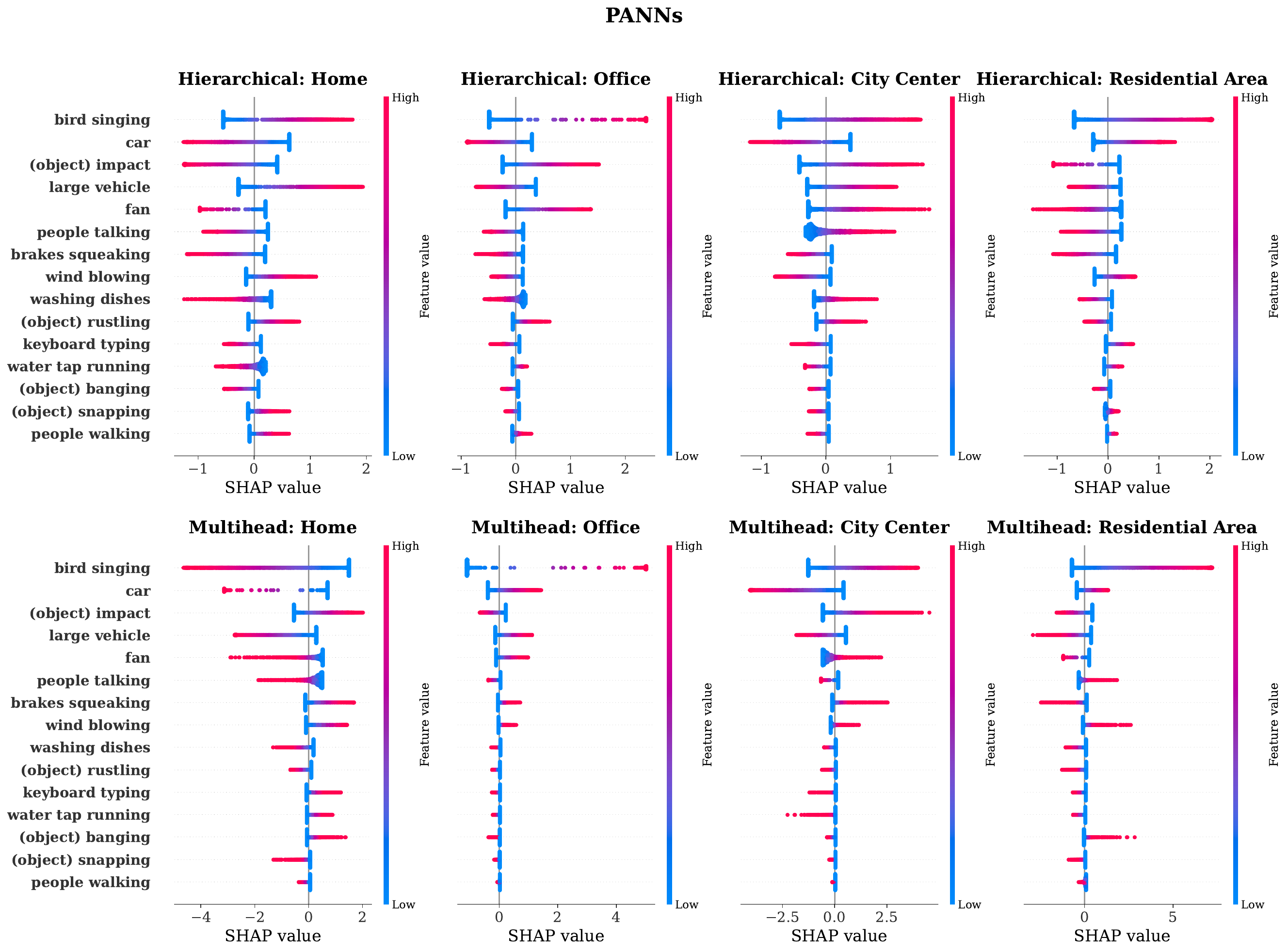}
    \caption{SHAP value summary plot comparing feature importance for PANNs under Multihead and Hierarchical configurations.}
    \label{fig:shap_panns}
\end{figure}

\begin{figure}[h]
    \centering
    \includegraphics[width=1\linewidth]{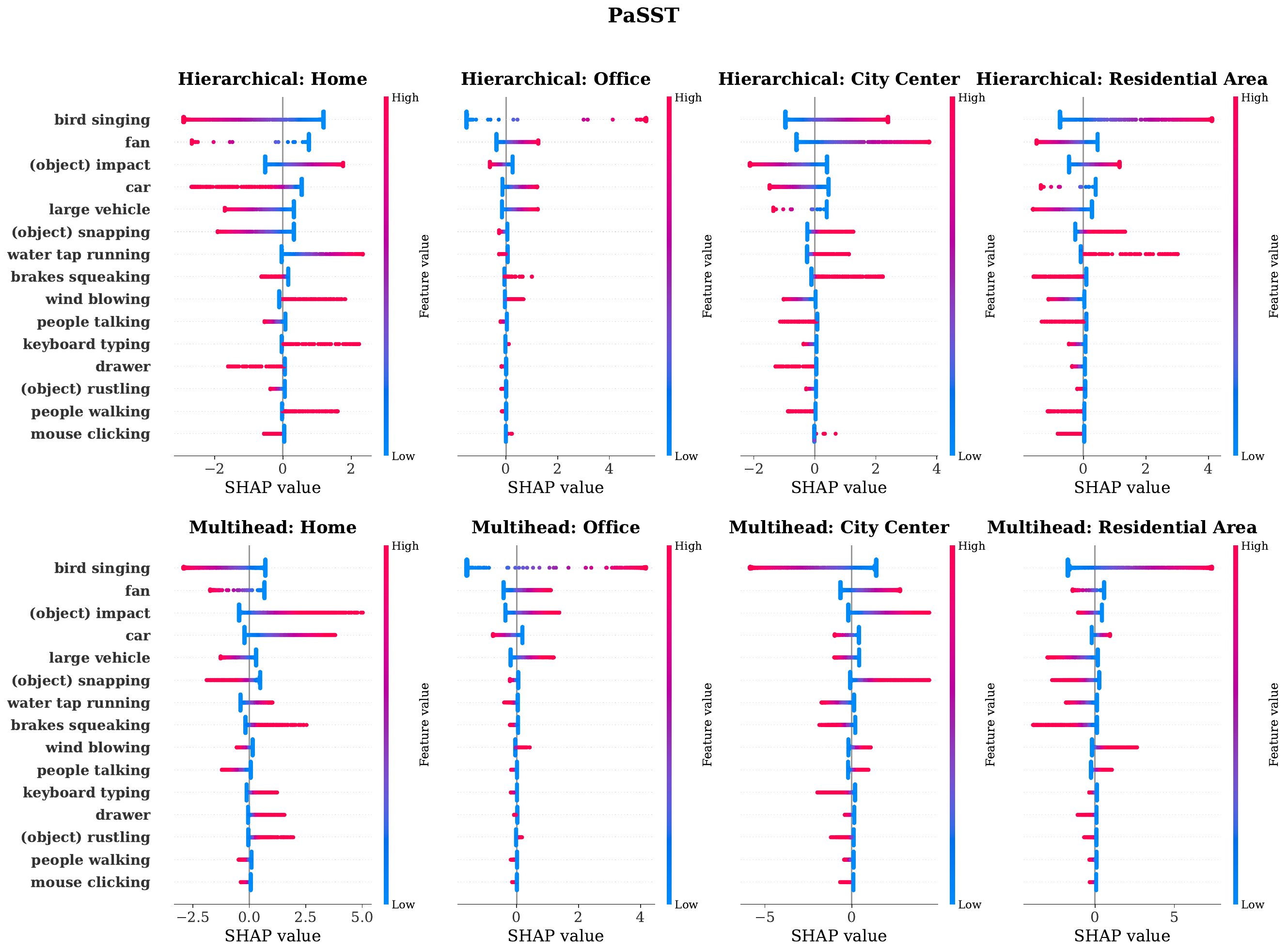}
    \caption{SHAP value summary plot comparing feature importance for PaSST under Multihead and Hierarchical configurations.}
    \label{fig:shap_passt}
\end{figure}

To complement the global information-theoretic and concept lattice analyses, we compute SHAP values to examine local feature importance at the event-to-scene prediction interface. Table \ref{tab:shap_scene_importance} illustrates the most important event for each scene, model, and architecture, and Figures \ref{fig:shap_beats}, \ref{fig:shap_panns}, and \ref{fig:shap_passt} illustrate the impact of individual intermediate acoustic event activation probabilities on final scene classification decisions across BEATs, PANNs, and PaSST, respectively. 

The most influential events are largely consistent across models and architectures. In particular, "car", "fan", and "bird singing" repeatedly emerge as the dominant events for "city center", "office", and "residential area", respectively, while the events associated with "home" show greater variation. These results provide a complementary view of the event-scene relationships identified by the FCA analysis, by highlighting the events that contribute most strongly to scene predictions.

\begin{table}[]
\centering
\small
\caption{Most influential acoustic event for each scene according to mean SHAP values.}
\label{tab:shap_scene_importance}
\begin{tabular}{cccc}
\toprule
\textbf{Model} & \textbf{Arch.} & \textbf{Scene} & \textbf{Top event} \\
\midrule
\multirow[c]{8}{*}{PaSST} & \multirow{4}{*}{Mult.} & home & water tap running (+1.60) \\
\cmidrule{3-4}
 &  & office & fan (+3.93) \\
\cmidrule{3-4}
 &  & city center & car (+1.59) \\
\cmidrule{3-4}
 &  & residential area & bird singing (+5.27) \\
\cmidrule{2-4}
 & \multirow{4}{*}{Hier.} & home & car (+1.18) \\
\cmidrule{3-4}
 &  & office & fan (+5.28) \\
\cmidrule{3-4}
 &  & city center & car (+1.46) \\
\cmidrule{3-4}
 &  & residential area & bird singing (+2.49) \\
\midrule
\multirow{8}{*}{PANNs} & \multirow{4}{*}{Mult.} & home & car (+1.47) \\
\cmidrule{3-4}
 &  & office & fan (+4.89) \\
\cmidrule{3-4}
 &  & city center & car (+2.39) \\
\cmidrule{3-4}
 &  & residential area & bird singing (+4.31) \\
\cmidrule{2-4}
 & \multirow{4}{*}{Hier.} & home & washing dishes (+1.27) \\
\cmidrule{3-4}
 &  & office & fan (+2.34) \\
\cmidrule{3-4}
 &  & city center & car (+0.91) \\
\cmidrule{3-4}
 &  & residential area & bird singing (+1.67) \\
\midrule
\multirow{8}{*}{BEATs} & \multirow{4}{*}{Mult.} & home & car (+1.39) \\
\cmidrule{3-4}
 &  & office & fan (+5.91) \\
\cmidrule{3-4}
 &  & city center & car (+1.50) \\
\cmidrule{3-4}
 &  & residential area & bird singing (+3.34) \\
\cmidrule{2-4}
 & \multirow{4}{*}{Hier.} & home & (object) impact (+1.14) \\
\cmidrule{3-4}
 &  & office & fan (+4.68) \\
\cmidrule{3-4}
 &  & city center & car (+1.42) \\
\cmidrule{3-4}
 &  & residential area & bird singing (+6.07) \\
\bottomrule
\end{tabular}
\end{table}

\subsection{Confussion Matrix and Co-occurence Matrix} \label{appendix:cm}

\begin{figure*}[t]
    \centering
    \includegraphics[width=1\textwidth]{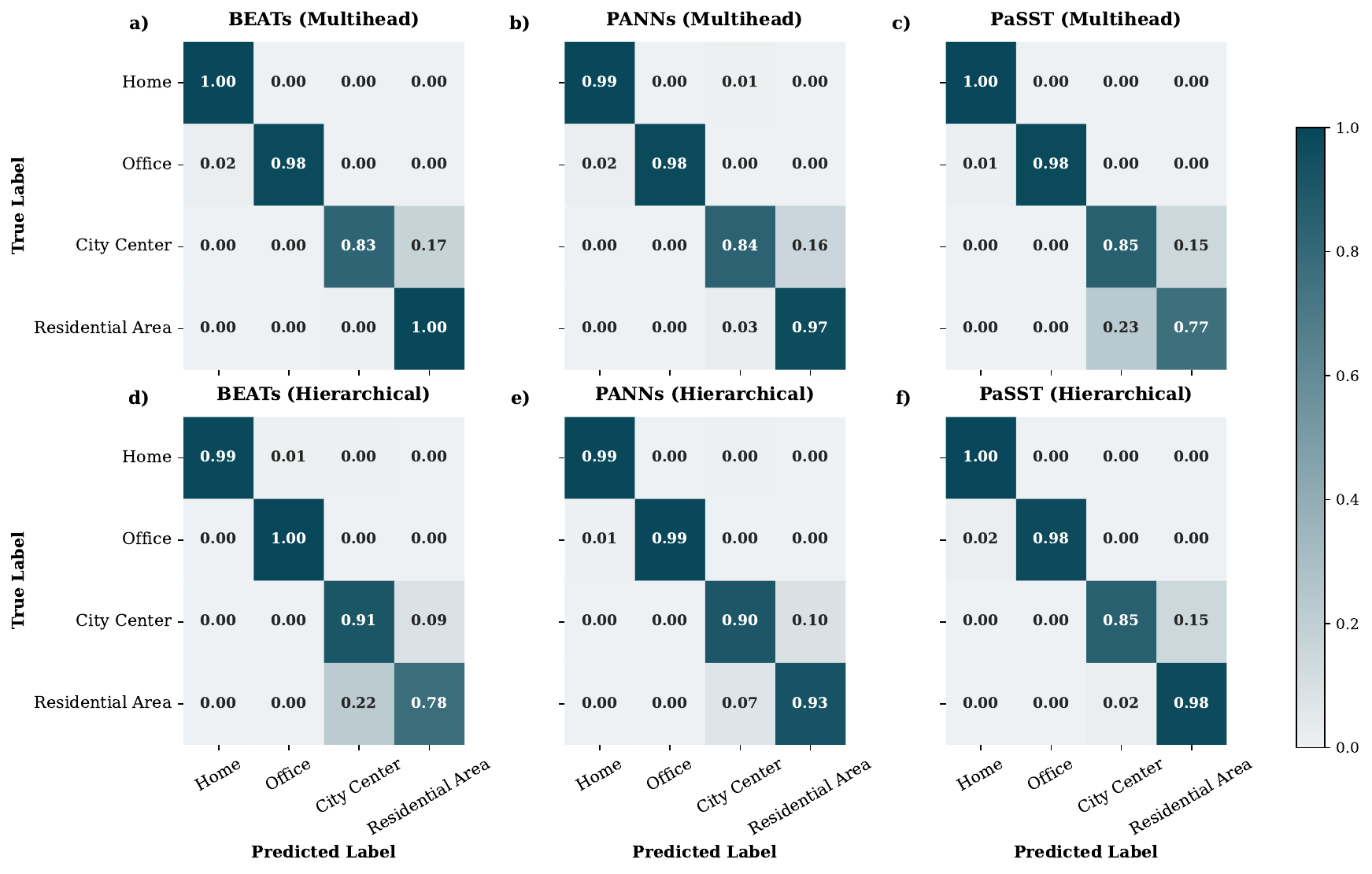}
    \caption{Normalized confusion matrices for ASC on the evaluation set. Each column corresponds to an acoustic model, namely \textbf{(a,d)} BEATs, \textbf{(b,e)} PANNs, and \textbf{(c,f)} PaSST, while the top and bottom rows \textbf{(a-c)} represent the multihead and \textbf{(d-f)} hierarchical architectures, respectively.}
    \label{fig:cm_ASC}
\end{figure*}

\begin{figure*}[b]
    \centering
    \includegraphics[width=1\textwidth]{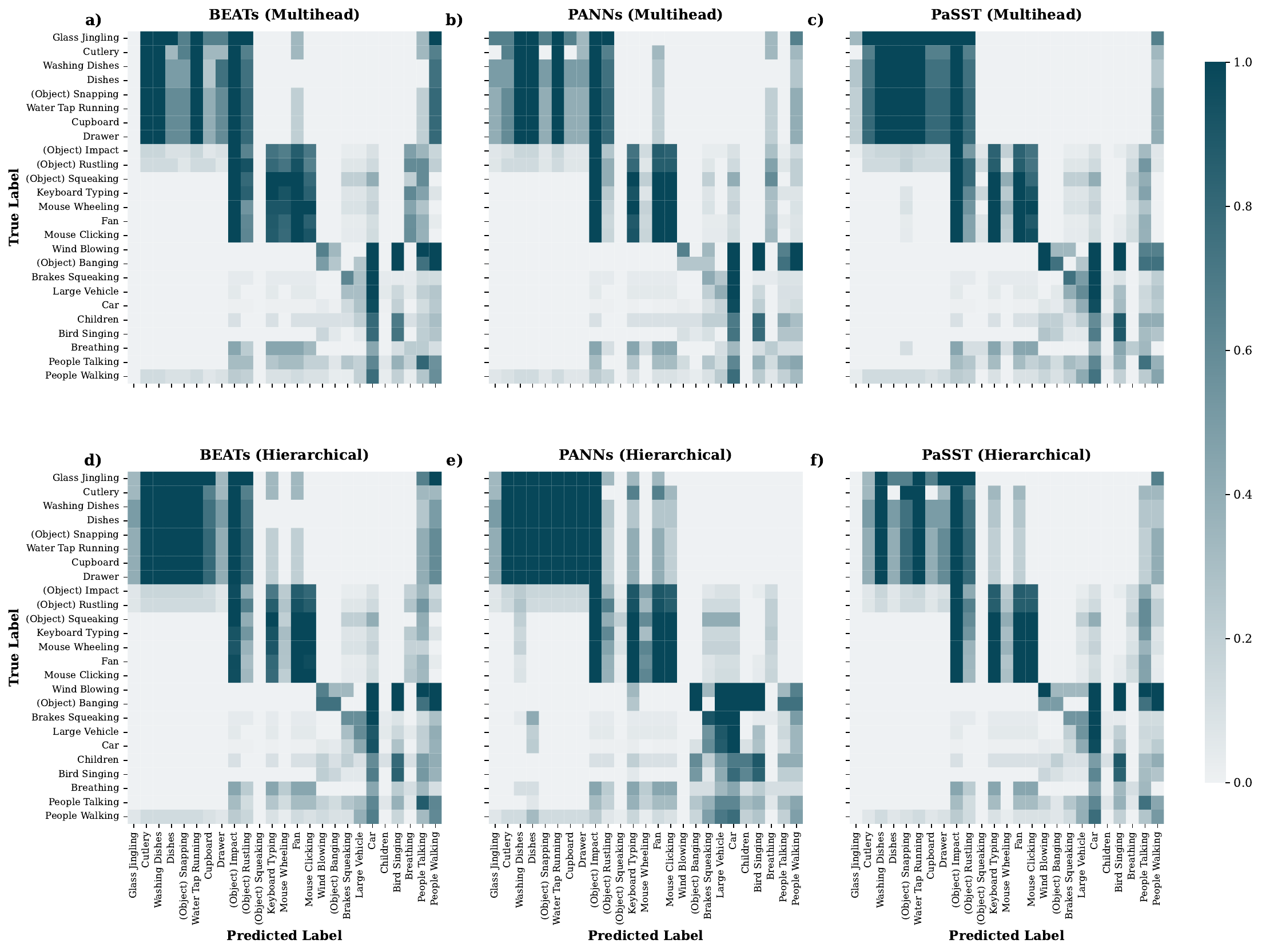}
    \caption{Normalized co-occurrence matrices for AT on the evaluation set. Each column corresponds to an acoustic model, namely \textbf{(a,d)} BEATs, \textbf{(b,e)} PANNs, and \textbf{(c,f)} PaSST, while the top and bottom rows \textbf{(a-c)} represent the multihead and \textbf{(d-f)} hierarchical architectures, respectively.}
    \label{fig:cm_AT}
\end{figure*}

Fig. \ref{fig:cm_ASC} shows the normalized confusion matrices for the ASC task, while Fig. \ref{fig:cm_AT} illustrates the co-occurrence matrix for AT. For ASC, the main confusion in predictions is between outdoor scenes, i.e., "residential area" and "city center". Additionally, in some cases, there is a little confusion between "home" and "office", but never between indoor and outdoor scenes. On the other hand, regarding performance in AT, it can be seen that there is a huge confusion between sounds belonging to a given acoustic scene, so we can see clusters in Fig. \ref{fig:cm_AT}. 

\end{document}